\input epsf
\input amssym
\input colordvi
\catcode`\@11\relax
\newif\ify@autoscale \y@autoscaletrue \def\Yautoscale#1{\ifnum #1=0
  \y@autoscalefalse\else\y@autoscaletrue\fi}
\newdimen\y@b@xdim
\newdimen\y@boxdim \y@boxdim=13pt
\def\Yboxdim#1{\y@autoscalefalse\y@boxdim=#1}
\newdimen\y@linethick    \y@linethick=.3pt
\def\Ylinethick#1{\y@linethick=#1}
\newskip\y@interspace \y@interspace=0ex plus 0.3ex
\def\Yinterspace#1{\y@interspace=#1}
\newif\ify@vcenter   \y@vcenterfalse
\def\Yvcentermath#1{\ifnum #1=0 \y@vcenterfalse\else\y@vcentertrue\fi}
\newif\ify@stdtext   \y@stdtextfalse
\def\Ystdtext#1{\ifnum #1=0 \y@stdtextfalse\else\y@stdtexttrue\fi}
\newif\ify@enable@skew   \y@enable@skewfalse
\expandafter\ifx\csname enableskew\endcsname\relax
 \y@enable@skewfalse \else \y@enable@skewtrue\fi
\def\y@vr{\vrule height0.8\y@b@xdim width\y@linethick depth 0.2\y@b@xdim}
\def\y@emptybox{\y@vr\hbox to \y@b@xdim{\hfil}}
\ify@enable@skew
 \def\y@abcbox#1{\if :#1\else
   \y@vr\hbox to \y@b@xdim{\hfil#1\hfil}\fi}
 \def\y@mathabcbox#1{\if :#1\else
   \y@vr\hbox to \y@b@xdim{\hfil$#1$\hfil}\fi}
\else
 \def\y@abcbox#1{\y@vr\hbox to \y@b@xdim{\hfil#1\hfil}}
 \def\y@mathabcbox#1{\y@vr\hbox to \y@b@xdim{\hfil$#1$\hfil}}
\fi
\def\y@setdim{%
  \ify@autoscale%
   \ifvoid1\else\typeout{Package youngtab: box1 not free! Expect an
     error!}\fi%
   \setbox1=\hbox{A}\y@b@xdim=1.6\ht1 \setbox1=\hbox{}\box1%
  \else\y@b@xdim=\y@boxdim \advance\y@b@xdim by -2\y@linethick
  \fi}
\newcount\y@counter
\newif\ify@islastarg
\def\y@lastargtest#1,#2 {\if\space #2 \y@islastargtrue
  \else\y@islastargfalse\fi}
\def\y@emptyboxes#1{\y@counter=#1\loop\ifnum\y@counter>0
  \advance\y@counter by -1 \y@emptybox\repeat}
\def\y@nelineemptyboxes#1{%
  \vbox{%
    \hrule height\y@linethick%
    \hbox{\y@emptyboxes{#1}\y@vr}
    \hrule height\y@linethick}\vskip-\y@linethick}
\def\yng(#1){%
  \y@setdim%
  \hskip\y@interspace%
  \ifmmode\ify@vcenter\vcenter\fi\fi{%
  \y@lastargtest#1,
  \vbox{\offinterlineskip
    \ify@islastarg
     \y@nelineemptyboxes{#1}
    \else
     \y@ungempty(#1)
    \fi}}\hskip\y@interspace}
\def\y@ungempty(#1,#2){%
  \y@nelineemptyboxes{#1}
  \y@lastargtest#2,
  \ify@islastarg
   \y@nelineemptyboxes{#2}
  \else
   \y@ungempty(#2)
  \fi}
\def\y@nelettertest#1#2. {\if\space #2 \y@islastargtrue
  \else\y@islastargfalse\fi}
\def\y@abcboxes#1#2.{%
  \ify@stdtext\y@abcbox#1\else\y@mathabcbox#1\fi%
  \y@nelettertest #2.
  \ify@islastarg\unskip%
   \ify@stdtext\y@abcbox{#2}\else\y@mathabcbox{#2}\fi%
  \else\y@abcboxes#2.\fi}
 \newdimen\y@full@b@xdim
 \newcount\y@m@veright@cnt
\ify@enable@skew
 \def\y@get@m@veright@cnt#1#2.{%
   \if :#1 \advance\y@m@veright@cnt by 1\y@get@m@veright@cnt#2.\fi}
 \let\y@setdim@=\y@setdim
 \def\y@setdim{%
   \y@setdim@ \y@full@b@xdim=\y@b@xdim
   \advance\y@full@b@xdim by 1\y@linethick}
 \def\y@m@veright@ifskew#1{
   \y@m@veright@cnt=0 \y@get@m@veright@cnt#1.
   \moveright \y@m@veright@cnt\y@full@b@xdim}
\else
 \def\y@m@veright@ifskew#1{}
\fi
\def\y@nelineabcboxes#1{%
  \y@nelettertest #1.
  \ify@islastarg
   \y@m@veright@ifskew{#1}
    \vbox{
      \hrule height\y@linethick%
      \hbox{\ify@stdtext\y@abcbox#1\else\y@mathabcbox#1\fi\y@vr}
      \hrule height\y@linethick}\vskip-\y@linethick
  \else
   \y@m@veright@ifskew{#1}
    \vbox{
      \hrule height\y@linethick%
      \hbox{\y@abcboxes #1.\y@vr}%
      \hrule height\y@linethick}\vskip-\y@linethick
  \fi}
\def\young(#1){%
  \y@setdim%
  \hskip\y@interspace%
  \y@lastargtest#1,
  \ifmmode\ify@vcenter\vcenter\fi\fi{%
  \vbox{\offinterlineskip
    \ify@islastarg\y@nelineabcboxes{#1}%
    \else\y@ungabc(#1)%
    \fi}}\hskip\y@interspace}
\def\y@ungabc(#1,#2){%
  \y@nelineabcboxes{#1}%
  \y@lastargtest#2,
  \ify@islastarg\y@nelineabcboxes{#2}%
  \else\y@ungabc(#2)%
  \fi}
\catcode`\@12\relax
 

\Yvcentermath1
\Yboxdim{4pt}
\newfam\scrfam
\batchmode\font\tenscr=rsfs10 \errorstopmode
\ifx\tenscr\nullfont
        \message{rsfs script font not available. Replacing with calligraphic.}
        \def\scr{\cal}
\else   
        \font\sevenscr=rsfs7
        \font\fivescr=rsfs5
        \skewchar\tenscr='177 \skewchar\sevenscr='177 \skewchar\fivescr='177
        \textfont\scrfam=\tenscr \scriptfont\scrfam=\sevenscr
        \scriptscriptfont\scrfam=\fivescr
        \def\scr{\fam\scrfam}
        \def\cal{\scr}
\fi
\catcode`\@=11
\newfam\frakfam
\batchmode\font\tenfrak=eufm10 \errorstopmode
\ifx\tenfrak\nullfont
        \message{eufm font not available. Replacing with italic.}
        \def\frak{\it}
\else
	
	\font\sevenfrak=eufm7 \font\fivefrak=eufm5
        
	\textfont\frakfam=\tenfrak
	\scriptfont\frakfam=\sevenfrak \scriptscriptfont\frakfam=\fivefrak
	\def\frak{\fam\frakfam}
\fi
\catcode`\@=\active
\newfam\msbfam
\batchmode\font\twelvemsb=msbm10 scaled\magstep1 \errorstopmode
\ifx\twelvemsb\nullfont\def\Bbb{\bf}
        
	\font\eightbbb=cmb10 at 8pt
	\message{Blackboard bold not available. Replacing with boldface.}
\else   \catcode`\@=11
        \font\tenmsb=msbm10 \font\sevenmsb=msbm7 \font\fivemsb=msbm5
        \textfont\msbfam=\tenmsb
        \scriptfont\msbfam=\sevenmsb \scriptscriptfont\msbfam=\fivemsb
        \def\Bbb{\relax\expandafter\Bbb@}
        \def\Bbb@#1{{\Bbb@@{#1}}}
        \def\Bbb@@#1{\fam\msbfam\relax#1}
        \catcode`\@=\active
	
	\font\eightbbb=msbm8
\fi
        \font\fivemi=cmmi5
        \font\sixmi=cmmi6
        \font\eightrm=cmr8              \def\xrm{\eightrm}
        \font\eightbf=cmbx8             \def\xbf{\eightbf}
        \font\eightit=cmti10 at 8pt     \def\xit{\eightit}
                
        \font\eighttt=cmtt8             
        \font\eightcp=cmcsc8
        \font\eighti=cmmi8              \def\xold{\eighti}
        \font\eightmi=cmmi8
        \font\eightib=cmmib8             \def\xbold{\eightib}
        \font\teni=cmmi10               \def\old{\teni}
        \font\tencp=cmcsc10

        \font\twelvecp=cmcsc10 scaled\magstep1
        
        \font\sixrm=cmr6
        \font\fiverm=cmr5

        \font\eightsy=cmsy8
        \font\sixsy=cmsy6
        \font\eightsl=cmsl8
        \font\sixbf=cmbx6

	 at10pt	
	\font\twelvehelvbold=phvb at12pt
	 at14pt
	\font\sixteenhelvbold=phvb at16pt
	 at16pt

\def\noblackbox{\overfullrule=0pt}
\noblackbox

\def\eightpoint{
\def\rm{\fam0\eightrm}
\textfont0=\eightrm \scriptfont0=\sixrm \scriptscriptfont0=\fiverm
\textfont1=\eightmi  \scriptfont1=\sixmi  \scriptscriptfont1=\fivemi
\textfont2=\eightsy \scriptfont2=\sixsy \scriptscriptfont2=\fivesy
\textfont3=\tenex   \scriptfont3=\tenex \scriptscriptfont3=\tenex
\textfont\itfam=\eightit \def\it{\fam\itfam\eightit}
\textfont\slfam=\eightsl \def\sl{\fam\slfam\eightsl}
\textfont\ttfam=\eighttt \def\tt{\fam\ttfam\eighttt}
\textfont\bffam=\eightbf \scriptfont\bffam=\sixbf 
                         \scriptscriptfont\bffam=\fivebf
                         \def\bf{\fam\bffam\eightbf}
\normalbaselineskip=10pt}

\newtoks\headtext
\headline={\ifnum\pageno=1\hfill\else
	\ifodd\pageno{\eightcp\the\headtext}{ }\dotfill{ }{\old\folio}
	\else{\old\folio}{ }\dotfill{ }{\eightcp\the\headtext}\fi
	\fi}
\def\makeheadline{\vbox to 0pt{\vss\noindent\the\headline\break
\hbox to\hsize{\hfill}}
        \vskip2\baselineskip}
\newcount\infootnote
\infootnote=0
\newcount\footnotecount
\footnotecount=1
\def\foot#1{\infootnote=1
\footnote{${}^{\the\footnotecount}$}{\vtop{\baselineskip=.75\baselineskip
\advance\hsize by
-\parindent{\eightpoint\rm\hskip-\parindent
#1}\hfill\vskip\parskip}}\infootnote=0\global\advance\footnotecount by
1}
\newcount\refcount
\refcount=1
\newwrite\refwrite
\def\oldsize{\ifnum\infootnote=1\xold\else\old\fi}
\def\ref#1#2{
	\def#1{{{\oldsize\the\refcount}}\ifnum\the\refcount=1\immediate\openout\refwrite=\jobname.refs\fi\immediate\write\refwrite{\item{[{\xold\the\refcount}]} 
	#2\hfill\par\vskip-2pt}\xdef#1{{\noexpand\oldsize\the\refcount}}\global\advance\refcount by 1}
	}
\def\refout{\eightpoint\catcode`\@=11
        \xrm\immediate\closeout\refwrite
        \vskip2\baselineskip
        {\noindent\twelvecp References}\hfill\vskip\baselineskip
        \baselineskip=.75\baselineskip
        \input\jobname.refs
        \baselineskip=4\baselineskip \divide\baselineskip by 3
        \catcode`\@=\active\rm}

\def\skipref#1{\hbox to15pt{\phantom{#1}\hfill}\hskip-15pt}

\def\hepth#1{\href{http://xxx.lanl.gov/abs/hep-th/#1}{arXiv:\allowbreak
hep-th\slash{\xold#1}}}

\def\arxiv#1#2{\href{http://arxiv.org/abs/#1.#2}{arXiv:\allowbreak
{\xold#1}.{\xold#2}}} 
 
\def\jhep#1#2#3#4{\href{http://jhep.sissa.it/stdsearch?paper=#2\%28#3\%29#4}{J. High Energy Phys. {\xbold #1#2} ({\xold#3}) {\xold#4}}}

\def\CQG#1#2#3{Class. Quantum Grav. {\xbold#1} ({\xold#2}) {\xold#3}}
\def\FP#1#2#3{Fortsch. Phys. {\xbold#1} ({\xold#2}) {\xold#3}}

\def\IJMPA#1#2#3{Int. J. Mod. Phys. {\xbf A}{\xbold#1} ({\xold#2}) {\xold#3}}

\def\JMP#1#2#3{J. Math. Phys. {\xbold#1} ({\xold#2}) {\xold#3}}
\def\JPA#1#2#3{J. Phys. {\xbf A}{\xbold#1} ({\xold#2}) {\xold#3}}

\def\NPB#1#2#3{Nucl. Phys. {\xbf B}{\xbold#1} ({\xold#2}) {\xold#3}}

\def\PLB#1#2#3{Phys. Lett. {\xbf B}{\xbold#1} ({\xold#2}) {\xold#3}}

\def\PRD#1#2#3{Phys. Rev. {\xbf D}{\xbold#1} ({\xold#2}) {\xold#3}}

\newcount\sectioncount
\sectioncount=0
\def\section#1#2{\global\eqcount=0
	\global\subsectioncount=0
        \global\advance\sectioncount by 1
	\ifnum\sectioncount>1
	        \vskip2\baselineskip
	\fi
\noindent{\twelvecp\the\sectioncount. #2}\par\nobreak
       \vskip.5\baselineskip\noindent
        \xdef#1{{\old\the\sectioncount}}}
\newcount\subsectioncount
\def\subsection#1#2{\global\advance\subsectioncount by 1
\vskip.75\baselineskip\noindent\line{\tencp\the\sectioncount.\the\subsectioncount. #2\hfill}\nobreak\vskip.4\baselineskip\nobreak\noindent\xdef#1{{\old\the\sectioncount}.{\old\the\subsectioncount}}}
\def\immediatesubsection#1#2{\global\advance\subsectioncount by 1
\vskip-\baselineskip\noindent
\line{\tencp\the\sectioncount.\the\subsectioncount. #2\hfill}
	\vskip.5\baselineskip\noindent
	\xdef#1{{\old\the\sectioncount}.{\old\the\subsectioncount}}}
\newcount\subsubsectioncount
\def\subsubsection#1#2{\global\advance\subsubsectioncount by 1
\vskip.75\baselineskip\noindent\line{\tencp\the\sectioncount.\the\subsectioncount.\the\subsubsectioncount. #2\hfill}\nobreak\vskip.4\baselineskip\nobreak\noindent\xdef#1{{\old\the\sectioncount}.{\old\the\subsectioncount}.{\old\the\subsubsectioncount}}}
\newcount\appendixcount
\appendixcount=0
\def\appendix#1{\global\eqcount=0
        \global\advance\appendixcount by 1
        \vskip2\baselineskip\noindent
        \ifnum\the\appendixcount=1
        {\twelvecp Appendix A: #1}\par\nobreak
                        \vskip.5\baselineskip\noindent\fi
        \ifnum\the\appendixcount=2
        {\twelvecp Appendix B: #1}\par\nobreak
                        \vskip.5\baselineskip\noindent\fi
        \ifnum\the\appendixcount=3
        {\twelvecp Appendix C: #1}\par\nobreak
                        \vskip.5\baselineskip\noindent\fi}
\def\acknowledgements{\vskip2\baselineskip\noindent
        \underbar{\it Acknowledgements:}\ }
\newcount\eqcount
\eqcount=0
\def\Eqn#1{\global\advance\eqcount by 1
\ifnum\the\sectioncount=0
	\xdef#1{{\noexpand\oldsize\the\eqcount}}
	\eqno({\oldstyle\the\eqcount})
\else
        \ifnum\the\appendixcount=0
\xdef#1{{\noexpand\oldsize\the\sectioncount}.{\noexpand\oldsize\the\eqcount}}
                \eqno({\oldstyle\the\sectioncount}.{\oldstyle\the\eqcount})\fi
        \ifnum\the\appendixcount=1
	        \xdef#1{{\noexpand\oldstyle A}.{\noexpand\oldstyle\the\eqcount}}
                \eqno({\oldstyle A}.{\oldstyle\the\eqcount})\fi
        \ifnum\the\appendixcount=2
	        \xdef#1{{\noexpand\oldstyle B}.{\noexpand\oldstyle\the\eqcount}}
                \eqno({\oldstyle B}.{\oldstyle\the\eqcount})\fi
        \ifnum\the\appendixcount=3
	        \xdef#1{{\noexpand\oldstyle C}.{\noexpand\oldstyle\the\eqcount}}
                \eqno({\oldstyle C}.{\oldstyle\the\eqcount})\fi
\fi}
\def\eqn{\global\advance\eqcount by 1
\ifnum\the\sectioncount=0
	\eqno({\oldstyle\the\eqcount})
\else
        \ifnum\the\appendixcount=0
                \eqno({\oldstyle\the\sectioncount}.{\oldstyle\the\eqcount})\fi
        \ifnum\the\appendixcount=1
                \eqno({\oldstyle A}.{\oldstyle\the\eqcount})\fi
        \ifnum\the\appendixcount=2
                \eqno({\oldstyle B}.{\oldstyle\the\eqcount})\fi
        \ifnum\the\appendixcount=3
                \eqno({\oldstyle C}.{\oldstyle\the\eqcount})\fi
\fi}
\def\multi{\global\advance\eqcount by 1}
\def\multieqn#1{({\oldstyle\the\sectioncount}.{\oldstyle\the\eqcount}\hbox{#1})}
\def\multiEqn#1#2{\xdef#1{{\oldstyle\the\sectioncount}.{\old\the\eqcount}#2}
        ({\oldstyle\the\sectioncount}.{\oldstyle\the\eqcount}\hbox{#2})}
\def\multiEqnAll#1{\xdef#1{{\oldstyle\the\sectioncount}.{\old\the\eqcount}}}
\newcount\tablecount
\tablecount=0
\def\Table#1#2{\global\advance\tablecount by 1
       \xdef#1{\the\tablecount}
       \vskip2\parskip
       \centerline{\it Table \the\tablecount: #2}
       \vskip2\parskip}
\newtoks\url
\def\Href#1#2{\catcode`\#=12\url={#1}\catcode`\#=\active#2}
\def\href#1#2{{#2}}

\parskip=3.5pt plus .3pt minus .3pt
\baselineskip=14pt plus .1pt minus .05pt
\lineskip=.5pt plus .05pt minus .05pt
\lineskiplimit=.5pt
\abovedisplayskip=18pt plus 4pt minus 2pt
\belowdisplayskip=\abovedisplayskip
\hsize=14cm
\vsize=19cm
\hoffset=1.5cm
\voffset=1.8cm
\frenchspacing
\footline={}
\raggedbottom

\newskip\origparindent
\origparindent=\parindent

\def\*{\partial}
\def\punkt{\,\,.}
\def\komma{\,\,,}

\def\={\!=\!}
\def\small#1{{\hbox{$#1$}}}

\def\fraction#1{\small{1\over#1}}
\def\fr{\fraction}
\def\Fraction#1#2{\small{#1\over#2}}
\def\Fr{\Fraction}
\def\tr{\hbox{\rm tr}}
\def\eg{{\it e.g.}}

\def\ie{{\it i.e.}}

\def\d{\delta}

\def\G{\Gamma}

\def\RR{{\Bbb R}}



\def\ol{\overline}

\def\textfrac#1#2{\raise .45ex\hbox{\the\scriptfont0 #1}\nobreak\hskip-1pt/\hskip-1pt\hbox{\the\scriptfont0 #2}}

\def\LL{{\cal L}}
\def\leftbr{[\![}
\def\rightbr{]\!]}
\def\leftpar{(\!(}
\def\rightpar{)\!)}


\def\frac{\Fr}

\def\mathbb{\Bbb}

\def\ZZ{{\Bbb Z}}



\def\LL{{\cal L}}
\def\leftbr{[\![}
\def\rightbr{]\!]}
\def\leftpar{(\!(}
\def\rightpar{)\!)}

\def\LL{{\cal L}}
\def\leftbr{[\![}
\def\rightbr{]\!]}
\def\leftpar{(\!(}
\def\rightpar{)\!)}

\def\ad{\hbox{ad}}

\def\LL{{\cal L}}

\def\leftbr{[\![}
\def\rightbr{]\!]}
\def\leftpar{(\!(}
\def\rightpar{)\!)}

\def\fe{{\frak e}}

\def\ol{\overline}

\def\ringaccent#1{{\mathaccent23 #1}}

\def\LLO{\ringaccent\LL}

\def\ad{\hbox{ad}\hskip1pt}


\catcode`@=11
\def\openupnormal{\afterassignment\@penupnormal\dimen@=}
\def\@penupnormal{\advance\normallineskip\dimen@
  \advance\normalbaselineskip\dimen@
  \advance\normallineskiplimit\dimen@}
\catcode`@=12

\def\EqMatrix{\let\quad\enspace\openupnormal6pt\matrix}

%
\line{
\epsfysize=18mm
\epsffile{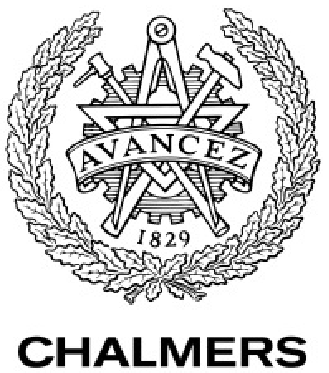}
\hskip3mm\epsfysize=18mm
\lower2pt\hbox{\epsffile{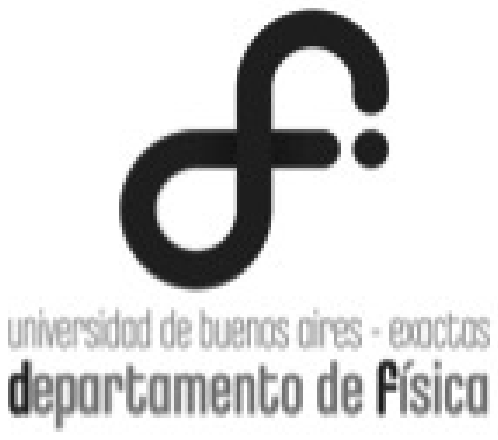}}
\hfill}
\vskip-11mm

\line{\hfill Gothenburg preprint}
\line{\hfill April, {\old2015}}

\line{\hrulefill}

\headtext={Cederwall, Rosabal: 
``{\eightmi E}\lower2pt\hbox{\sixmi 8} geometry''}

\vfill
\vskip.5cm

\centerline{\sixteenhelvbold
{\sixteenhelvbold E}\lower4pt\hbox{\twelvehelvbold 8} geometry}

\vfill

\centerline{\twelvehelvbold Martin Cederwall}

\vfill
\vskip-1cm

\centerline{\it Dept. of Fundamental Physics}
\centerline{\it Chalmers University of Technology}
\centerline{\it SE 412 96 Gothenburg, Sweden}

\vfill\vskip-.5cm

\centerline{\twelvehelvbold J.A. Rosabal}

\vfill
\vskip-1cm

\centerline{\it Departamento de F\'\i sica}
\centerline{\it Universidad de Buenos Aires CONICET-UBA}
\centerline{\it Pabell\'on I, Ciudad Universitaria, Buenos Aires, Argentina}

\vfill

{\narrower\noindent \underbar{Abstract:}
We investigate exceptional generalised diffeomorphisms based on
$E_{8(8)}$ in a geometric setting. 
The transformations include gauge transformations for the dual gravity field.
The surprising key result, which
allows for a development of a tensor formalism, is that it is possible
to define field-dependent transformations containing connection, which
are covariant. We solve for the spin connection and construct a
curvature tensor. A geometry for the Ehlers symmetry $SL(n+1)$ is
sketched. Some related issues are discussed.
\smallskip}
\vfill

\font\xxtt=cmtt6

\vtop{\baselineskip=.6\baselineskip\xxtt
\line{\hrulefill}
\catcode`\@=11
\line{email: martin.cederwall@chalmers.se, arosabal@df.uba.ar\hfill}
\catcode`\@=\active
}

\eject

%

\def\textfrac#1#2{\raise .45ex\hbox{\the\scriptfont0 #1}\nobreak\hskip-1pt/\hskip-1pt\hbox{\the\scriptfont0 #2}}

\def\fe{{\frak e}}

\def\boxit#1{\vbox{\hrule\hbox{\vrule\kern3pt
             \vbox{\kern3pt#1\kern3pt}\kern3pt\vrule}\hrule}}


\ref\Rosabal{J.A. Rosabal, {\xit ``On the exceptional generalised Lie
derivative for $d\geq7$''}, \arxiv{1410}{8148}.}

\ref\HohmSamtleben{O. Hohm and H. Samtleben, {\xit ``Exceptional field
theory III: $E_{8(8)}$''}, \PRD{90}{2014}{066002} [\arxiv{1406}{3348}].}

\ref\CederwallEdlundKarlsson{M. Cederwall, J. Edlund and A. Karlsson,
  {\xit ``Exceptional geometry and tensor fields''}, 
\jhep{13}{07}{2013}{028} [\arxiv{1302}{6736}].}

\ref\GodazgarGodazgarNicolai{H. Godazgar, M. Godazgar and H. Nicolai,
{\xit ``Einstein-Cartan calculus for exceptional geometry''},
\jhep{14}{06}{2014}{021} [\arxiv{1401}{5984}].}

\ref\BermanCederwallKleinschmidtThompson{D.S. Berman, M. Cederwall,
A. Kleinschmidt and D.C. Thompson, {\xit ``The gauge structure of
generalised diffeomorphisms''}, \jhep{13}{01}{2013}{64} [\arxiv{1208}{5884}].}

\ref\GodazgarGodazgarPerry{H. Godazgar, M. Godazgar and M.J. Perry,
{\xit ``$E_8$ duality and dual gravity''}, \jhep{13}{06}{2013}{044}
[\arxiv{1303}{2035}].} 

\ref\BermanCederwallPerry{D.S. Berman, M. Cederwall and M.J. Perry,
{\xit ``Global aspects of double geometry''}, 
\jhep{14}{09}{2014}{66} [\arxiv{1401}{1311}].}

\ref\BekaertBoulangerHenneaux{X. Bekaert, N. Boulanger, and
M. Henneaux, {\xit ``Consistent deformations of dual formulations of
linearized gravity: a no go result''}, \PRD{67}{2003}{044010}
[\hepth{0210278}].} 

\ref\deWitNicolaiSamtleben{B. de Wit, H. Nicolai and H. Samtleben,
{\xit ``Gauged supergravities, tensor hierarchies, and M-theory''},
\jhep{02}{08}{2008}{044} [\arxiv{0801}{1294}].}

\ref\Duff{M.J. Duff, {\xit ``Duality rotations in string
theory''}, \NPB{335}{1990}{610}.}

\ref\Tseytlin{A.A.~Tseytlin,
  {\xit ``Duality symmetric closed string theory and interacting
  chiral scalars''}, 
  \NPB{350}{1991}{395}.}

\ref\SiegelI{W.~Siegel,
  {\xit ``Two vierbein formalism for string inspired axionic gravity''},
  \PRD{47}{1993}{5453} [\hepth{9302036}].}

\ref\SiegelII{ W.~Siegel,
  {\xit ``Superspace duality in low-energy superstrings''},
  \PRD{48}{1993}{2826} [\hepth{9305073}].}

\ref\SiegelIII{W.~Siegel,
  {\xit ``Manifest duality in low-energy superstrings''},
  in Berkeley 1993, Proceedings, Strings '93 353
  [\hepth{9308133}].}

\ref\HullDoubled{C.M. Hull, {\xit ``Doubled geometry and
T-folds''}, \jhep{07}{07}{2007}{080} [\hepth{0605149}].}

\ref\HullT{C.M. Hull, {\xit ``A geometry for non-geometric string
backgrounds''}, \jhep{05}{10}{2005}{065} [\hepth{0406102}].}

\ref\HullM{C.M. Hull, {\xit ``Generalised geometry for M-theory''},
\jhep{07}{07}{2007}{079} [\hepth{0701203}].}

\ref\HullZwiebachDFT{C. Hull and B. Zwiebach, {\xit ``Double field
theory''}, \jhep{09}{09}{2009}{99} [\arxiv{0904}{4664}].}

\ref\HohmHullZwiebachI{O. Hohm, C.M. Hull and B. Zwiebach, {\xit ``Background
independent action for double field
theory''}, \jhep{10}{07}{2010}{016} [\arxiv{1003}{5027}].}

\ref\HohmHullZwiebachII{O. Hohm, C.M. Hull and B. Zwiebach, {\xit
``Generalized metric formulation of double field theory''},
\jhep{10}{08}{2010}{008} [\arxiv{1006}{4823}].} 

\ref\HohmKwak{O. Hohm and S.K. Kwak, {\xit ``$N=1$ supersymmetric
double field theory''}, \jhep{12}{03}{2012}{080} [\arxiv{1111}{7293}].}

\ref\HohmKwakFrame{O. Hohm and S.K. Kwak, {\xit ``Frame-like geometry
of double field theory''}, \JPA{44}{2011}{085404} [\arxiv{1011}{4101}].}

\ref\JeonLeeParkI{I. Jeon, K. Lee and J.-H. Park, {\xit ``Differential
geometry with a projection: Application to double field theory''},
\jhep{11}{04}{2011}{014} [\arxiv{1011}{1324}].}

\ref\JeonLeeParkII{I. Jeon, K. Lee and J.-H. Park, {\xit ``Stringy
differential geometry, beyond Riemann''}, 
\PRD{84}{2011}{044022} [\arxiv{1105}{6294}].}

\ref\JeonLeeParkIII{I. Jeon, K. Lee and J.-H. Park, {\xit
``Supersymmetric double field theory: stringy reformulation of supergravity''},
\PRD{85}{2012}{081501} [\arxiv{1112}{0069}].}

\ref\HohmZwiebachLarge{O. Hohm and B. Zwiebach, {\xit ``Large gauge
transformations in double field theory''}, \jhep{13}{02}{2013}{075}
[\arxiv{1207}{4198}].} 

\ref\Park{J.-H.~Park,
  {\xit ``Comments on double field theory and diffeomorphisms''},
  \jhep{13}{06}{2013}{098} [\arxiv{1304}{5946}].}

\ref\BermanCederwallPerry{D.S. Berman, M. Cederwall and M.J. Perry,
{\xit ``Global aspects of double geometry''}, 
\jhep{14}{09}{2014}{66} [\arxiv{1401}{1311}].}

\ref\CederwallGeometryBehind{M. Cederwall, {\xit ``The geometry behind
double geometry''}, 
\jhep{14}{09}{2014}{70} [\arxiv{1402}{2513}].}

\ref\HohmLustZwiebach{O. Hohm, D. L\"ust and B. Zwiebach, {\xit ``The
spacetime of double field theory: Review, remarks and outlook''},
\FP{61}{2013}{926} [\arxiv{1309}{2977}].} 

\ref\Papadopoulos{G. Papadopoulos, {\xit ``Seeking the balance:
Patching double and exceptional field theories''}, 
\jhep{14}{10}{2014}{089} [\arxiv{1402}{2586}].}

\ref\HullGlobal{C.M. Hull, 	
{\xit ``Finite gauge transformations and geometry in double field
theory''}, \arxiv{1406}{7794}.}

\ref\PachecoWaldram{P.P. Pacheco and D. Waldram, {\xit ``M-theory,
exceptional generalised geometry and superpotentials''},
\jhep{08}{09}{2008}{123} [\arxiv{0804}{1362}].}

\ref\Hillmann{C. Hillmann, {\xit ``Generalized $E_{7(7)}$ coset
dynamics and $D=11$ supergravity''}, \jhep{09}{03}{2009}{135}
[\arxiv{0901}{1581}].}

\ref\BermanPerryGen{D.S. Berman and M.J. Perry, {\xit ``Generalised
geometry and M-theory''}, \jhep{11}{06}{2011}{074} [\arxiv{1008}{1763}].}    

\ref\BermanGodazgarPerry{D.S. Berman, H. Godazgar and M.J. Perry,
{\xit ``SO(5,5) duality in M-theory and generalized geometry''},
\PLB{700}{2011}{65} [\arxiv{1103}{5733}].} 

\ref\BermanMusaevPerry{D.S. Berman, E.T. Musaev and M.J. Perry,
{\xit ``Boundary terms in generalized geometry and doubled field theory''},
\PLB{706}{2011}{228} [\arxiv{1110}{97}].} 

\ref\BermanGodazgarGodazgarPerry{D.S. Berman, H. Godazgar, M. Godazgar  
and M.J. Perry,
{\xit ``The local symmetries of M-theory and their formulation in
generalised geometry''}, \jhep{12}{01}{2012}{012}
[\arxiv{1110}{3930}].} 

\ref\BermanGodazgarPerryWest{D.S. Berman, H. Godazgar, M.J. Perry and
P. West,
{\xit ``Duality invariant actions and generalised geometry''}, 
\jhep{12}{02}{2012}{108} [\arxiv{1111}{0459}].} 

\ref\CoimbraStricklandWaldram{A. Coimbra, C. Strickland-Constable and
D. Waldram, {\xit ``$E_{d(d)}\times\hbox{\eightbbb R}^+$ generalised geometry,
connections and M theory'' }, \jhep{14}{02}{2014}{054} [\arxiv{1112}{3989}].} 

\ref\CoimbraStricklandWaldramII{A. Coimbra, C. Strickland-Constable and
D. Waldram, {\xit ``Supergravity as generalised geometry II:
$E_{d(d)}\times\hbox{\eightbbb R}^+$ and M theory''}, 
\jhep{14}{03}{2014}{019} [\arxiv{1212}{1586}].}  

\ref\BermanCederwallKleinschmidtThompson{D.S. Berman, M. Cederwall,
A. Kleinschmidt and D.C. Thompson, {\xit ``The gauge structure of
generalised diffeomorphisms''}, \jhep{13}{01}{2013}{64} [\arxiv{1208}{5884}].}

\ref\ParkSuh{J.-H. Park and Y. Suh, {\xit ``U-geometry: SL(5)''},
\jhep{14}{06}{2014}{102} [\arxiv{1302}{1652}].} 

\ref\CederwallII{ M.~Cederwall,
  {\xit ``Non-gravitational exceptional supermultiplets''},
  \jhep{13}{07}{2013}{025} [\arxiv{1302}{6737}].}

\ref\SambtlebenHohmI{O.~Hohm and H.~Samtleben,
  {\xit ``Exceptional field theory I: $E_{6(6)}$ covariant form of
  M-theory and type IIB''}, 
  \PRD{89}{2014}{066016} [\arxiv{1312}{0614}].}

\ref\SambtlebenHohmII{O.~Hohm and H.~Samtleben,
  {\xit ``Exceptional field theory II: $E_{7(7)}$''},
  \PRD{89}{2014}{066016} [\arxiv{1312}{4542}].}

\ref\CederwallUfoldBranes{M. Cederwall, {\xit ``M-branes on U-folds''},
in proceedings of 7th International Workshop ``Supersymmetries and
Quantum Symmetries'' Dubna, 2007 [\arxiv{0712}{4287}].}

\ref\HasslerLust{F. Hassler and D. L\"ust, {\xit ``Consistent
compactification of double field theory on non-geometric flux
backgrounds''}, \jhep{14}{05}{2014}{085} [\arxiv{1401}{5068}].}

\ref\CederwallDuality{M. Cederwall, {\xit ``T-duality and
non-geometric solutions from double geometry''}, \FP{62}{2014}{942}
[\arxiv{1409}{4463}].} 

\ref\AldazabalGranaMarquesRosabal{G. Aldazabal, M. Gra\~na,
D. Marqu\'es and J.A. Rosabal, {\xit ``Extended geometry and gauged
maximal supergravity''}, 
\jhep{13}{06}{2013}{046} [\arxiv{1302}{5419}].}

\ref\AldazabalGranaMarquesRosabalII{G. Aldazabal, M. Gra\~na,
D. Marqu\'es and J.A. Rosabal, {\xit ``The gauge structure of
exceptional field theories and the tensor hierarchy ''}, 
\jhep{14}{04}{2014}{049} [\arxiv{1312}{4549}].}

\ref\PalmkvistHierarchy{J. Palmkvist, {\xit ``Tensor hierarchies,
Borcherds algebras and $E_{11}$''}, \jhep{12}{02}{2012}{066}
[\arxiv{1110}{4892}].} 

\ref\GreitzHowePalmkvist{J. Greitz, P.S. Howe and J. Palmkvist, {\xit ``The
tensor hierarchy simplified''}, \CQG{31}{2014}{087001} [\arxiv{1308}{4972}].} 

\ref\PalmkvistTensor{J. Palmkvist, {\xit ``The tensor hierarchy
algebra''}, \JMP{55}{2014}{011701} [\arxiv{1305}{0018}].}

\ref\HowePalmkvist{P.S. Howe and J. Palmkvist, {\xit ``Forms and
algebras in (half-)maximal supergravity theories''}, 
\hfill\break\arxiv{1503}{00015}.}

\ref\CederwallPalmkvistBorcherds{M. Cederwall and J. Palmkvist, {\xit
``Superalgebras, constraints and partition functions''}, \arxiv{1503}{06215}.}

\ref\BermanBlairMalekPerry{D.S. Berman, C.D.A. Blair, E. Malek and
M.J. Perry, {\xit ``The $O_{D,D}$ geometry of string theory''},
\IJMPA{29}{2014}{1450080} [\arxiv{1303}{6727}].}

\ref\BlairMalek{C.D.A. Blair and E. Malek, {\xit ``Geometry and fluxes
of SL(5) exceptional field theory''}, \jhep{15}{03}{2015}{144} [\arxiv{1412}{0635}].}

\ref\BlumenhagenHasslerLust{R. Blumenhagen, F. Hassler and D. L\"ust,
{\xit ``Double field theory on group manifolds''},
\jhep{15}{02}{2015}{001} [\arxiv{1410}{6374}].}

\ref\BlumenhagenBosqueHasslerLust{R. Blumenhagen, P. du Bosque,
F. Hassler and D. L\"ust, 
{\xit ``Generalized metric formulation of double field theory on group
manifolds''}, \arxiv{1502}{02428}.}

\ref\ParkSuh{J.-H. Park and Y. Suh, {\xit ``U-geometry: SL(5)''}, 
\jhep{14}{06}{2014}{102} [\arxiv{1302}{1652}].}

\ref\KoepsellNicolaiSamtleben{K. Koepsell, H. Nicolai and
H. Samtleben, {\xit ``On the Yangian $[Y(e_8)]$ quantum symmetry of
maximal supergravity in two dimensions''}, \jhep{99}{04}{1999}{023}
[\hepth{9903111}].}

\ref\CoimbraStricklandWaldramTypeII{A. Coimbra, C. Strickland-Constable and
D. Waldram, {\xit ``Supergravity as generalised geometry I: Type II
theories''}, \jhep{11}{11}{2011}{091} [\arxiv{1107}{1733}].} 

\ref\StricklandConstable{C. Strickland-Constable, {\xit ``Subsectors,
Dynkin diagrams and new generalised geometries''}, \arxiv{1310}{4196}.}


\section\Introduction{Introduction}Doubled geometry and exceptional geometry
provide a means
to include all massless gauge fields in string theory or M-theory into
a unified setting, providing a geometric origin of T-duality
[\Duff\skipref\Tseytlin\skipref\SiegelI\skipref\SiegelII\skipref\SiegelIII\skipref\HullT\skipref\HullDoubled\skipref\HullZwiebachDFT\skipref\HohmHullZwiebachI\skipref\HohmHullZwiebachII\skipref\CoimbraStricklandWaldramTypeII\skipref\HohmKwakFrame\skipref\HohmKwak\skipref\JeonLeeParkI\skipref\JeonLeeParkII\skipref\JeonLeeParkIII\skipref\HohmZwiebachLarge\skipref\Park\skipref\BermanCederwallPerry\skipref\CederwallGeometryBehind\skipref\HohmLustZwiebach\skipref\Papadopoulos\skipref\HullGlobal\skipref\CederwallDuality\skipref\BlumenhagenHasslerLust-\BlumenhagenBosqueHasslerLust]
or U-duality
[\HullM\skipref\PachecoWaldram\skipref\Hillmann\skipref\BermanPerryGen\skipref\BermanGodazgarPerry\skipref\BermanGodazgarGodazgarPerry\skipref\BermanGodazgarPerryWest\skipref\CoimbraStricklandWaldram\skipref\CoimbraStricklandWaldramII\skipref\BermanCederwallKleinschmidtThompson\skipref\ParkSuh\skipref\CederwallEdlundKarlsson\skipref\CederwallII\skipref\CederwallUfoldBranes\skipref\AldazabalGranaMarquesRosabal\skipref\AldazabalGranaMarquesRosabalII\skipref\SambtlebenHohmI-\SambtlebenHohmII].

The purpose of the present paper is to extend the concept of extended
geometry to $E_8$, the U-duality group obtained when M-theory is
dimensionally reduced to 3 dimensions. We will however focus on the
geometric picture for the ``internal'' dimensions.
Some work on this case has been done previously. In
refs. [\CoimbraStricklandWaldram,\BermanCederwallKleinschmidtThompson], it was 
noted that a na\"\i ve attempt to extend the definition of generalised
diffeomorphisms fail to close --- the commutator of two such
transformations produce a local $E_8$ transformation of a restricted
kind. Hohm and Samtleben [\HohmSamtleben] 
nevertheless managed to base a description of 11-dimensional supergravity
in a 3+8 split on such transformations, however with the drawback that
a geometric understanding was lacking. It was observed by one of the
present authors [\Rosabal] that the form of the ``extra'' $E_8$ transformations
suggests an interpretation in terms of a connection. We will build on
the latter observation, and develop an $E_8$ geometry. It essentially
vindicates the conclusions in ref. [\HohmSamtleben].


The difficulty with $E_8$ is sometimes attributed to the occurrence of
a dual gravity field.
Constructing a geometry for $E_8$ may be a first step towards
incorporating dual gravity. If contact is to be made with the
infinite-dimensional cases of $E_9$, $E_{10}$ (and maybe $E_{11}$),
this is essential, especially since there are no-go theorems to
circumvent [\BekaertBoulangerHenneaux]. 
We will comment more on this in the discussion section.
A solution to the problem is also relevant for lower $n$, where
non-covariance occurs not at the level of the algebra of generalised
diffeomorphisms, but higher in their reducibility, where mixed
symmetry fields arise.

The paper is organised as follows: In section {\old2}, we discuss
exceptional geometry (the arguments are valid also for ordinary and
doubled geometry) from the perspective of covariance and closure. This
helps us, in section {\old3}, to get a better geometric understanding of
what happens for $E_8$, and leads us to a candidate field-dependent
transformation. In section {\old4}, it is shown that this transformation,
quite surprisingly, has
the required covariance property, and the algebra is examined.
Section {\old5} is devoted to the development of the geometric framework of
models based on this symmetry. We define torsion, solve for the spin
connection and find a Ricci scalar. Section {\old6} deals with reducibility
and covariance, for the $E_8$ case, but also for lower $n$. Section
{\old7} sketches the situation for the simpler 
case of Ehlers symmetry, where the dual gravity field also is
present. We end with a summary and discussion. 

\section\CovarianceAndClosure{Covariance and closure for generalised
diffeomorphisms}In this preparatory section, we will revisit the
concepts of covariance and closure, especially how they are linked
together, for the cases known to work: ordinary diffeomorphisms,
double diffeomorphisms, and exceptional diffeomorphisms for
$n\leq7$. 
The exceptional cases, of which $n=8$ is continuing the series, will
however be our model examples.
This will give us tools to use when analysing the case $n=8$.

Consider some generalised diffeomorphism, which is 
generated by $\LLO_\xi$
constructed with naked derivatives.
Let its action on a vector in the module $R_1$ of $E_{n(n)}$ be
defined by 
$$
\eqalign{
\LLO_\xi V^M&=L_\xi V^M+Y^{MN}{}_{PQ}\*_N\xi^PV^Q\cr
&=\xi^N\*_NV^M+Z^{MN}{}_{PQ}\*_N\xi^PV^Q\komma\cr
}\eqn
$$
where the $Z$ in the second
term ensures that the indices ${}^M{}_Q$ are projected on ${\frak
e}_{n(n)}\oplus\RR\subset{\frak gl}(|R_1|)$
[\CoimbraStricklandWaldram,\BermanCederwallKleinschmidtThompson]. 
Of course the expression applies also for ordinary diffeomorphisms and
for double diffeomorphisms.
The invariant tensors $Y$ or $Z$ for the exceptional series have been given
in diverse papers, \eg\ ref. [\BermanCederwallKleinschmidtThompson],
and will not be repeated here.
They satisfy the important identity
$$
\left(Y^{MN}{}_{TQ}Y^{TP}{}_{RS}-Y^{MN}{}_{RS}\delta^P_Q\right)
\*_{(N}\otimes\*_{P)}=0\komma\Eqn\NonLinOne
$$
which can equivalently be written
$$
\left(Z^{MN}{}_{TQ}Z^{TP}{}_{RS}+Z^{MP}{}_{RQ}\d^N_S\right)
\*_{(N}\otimes\*_{P)}=0\punkt\Eqn\NonLinTwo
$$
The tensor $Y$ governs the section condition $(\*\otimes\*)|_{\ol
R_2}=0$, which reads
$$
Y^{MN}{}_{PQ}\*_M\otimes\*_N=0\punkt\eqn
$$
While eq. (\NonLinOne) manifests the $R_2$ and $\ol R_2$
projections of the index pairs ${}^{MN}$ and ${}_{RS}$, the form (\NonLinTwo)
manifests the $\fe_8\oplus\RR$ projections in the pairs ${}^M{}_Q$ and
${}^P{}_R$.

There is a close connection between covariance and closure of the
algebra, and the former may be used to prove the latter. Let us first formalise
what covariance and closure means. The latter is simple, it means that
the generators commute to a transformation with some parameter:
$$
[\LLO_\xi,\LLO_\eta]V=\LLO_{\leftbr\xi,\eta\rightbr}V\komma\eqn
$$
where $\leftbr\cdot,\cdot\rightbr$ for the moment is an unspecified
bracket encoding the structure constants.
Covariance means, on the other hand, that the transformed vector
$\LLO_\xi V$ is a vector, when the vectorial transformation of both
$V$ and $\xi$ are taken into account. This may be written
$$
\hat\delta_\eta(\LLO_\xi V)
\equiv\LLO_\xi\LLO_\eta V+\LLO_{\LLO_\eta\xi}V=\LLO_\eta\LLO_\xi V
\Eqn\CovarianceEquation
$$
(the convention is that ``$\hat\delta$'' is used when also parameters
transform, unlike ``$\delta$'', which only transforms fields).
Assuming covariance immediately means that
$$
[\LLO_\xi,\LLO_\eta]V=-\LLO_{\LLO_\eta\xi}V\punkt\eqn
$$
This implies that the algebra closes, with 
$\leftbr\cdot,\cdot\rightbr=\fr2(\LLO_\xi\eta-\LLO_\eta\xi)$. In
addition, the left hand side is antisymmetric, so one also gets
$$
\LLO_{\leftpar\xi,\eta\rightpar}V=0\komma\Eqn\SymmZero
$$
where $\leftpar\cdot,\cdot\rightpar=\fr2(\LLO_\xi\eta+\LLO_\eta\xi)$.
Ordinary diffeomorphisms of course already have 
$\leftpar\xi,\eta\rightpar=0$, but for generalised diffeomorphisms for
$O(d,d)$ and $E_{n(n)}\times\RR^+$, $n\leq7$, eq. (\SymmZero) is
non-trivially satisfied. 
The covariance equation (\CovarianceEquation) can be used to show that
the Jacobiator $\leftbr\xi,\eta,\zeta\rightbr
\equiv\leftbr\xi,\leftbr\eta,\zeta\rightbr\rightbr+\hbox{\it cycl}$ then is non-zero,
but equal to such a null parameter:
$$
\leftbr\xi,\eta,\zeta\rightbr
=-\fr3\leftpar\xi,\leftbr\eta,\zeta\rightbr\rightpar+\hbox{\it cycl}\punkt
\Eqn\Jacobiator
$$ 

Thus, checking covariance is enough to ensure closure. If the naked
derivatives in $\LLO$ are replaced by covariant derivatives,
covariance becomes manifest. To show covariance therefore amounts to
demonstrating the absence of (the non-torsion part of) the connection
in 
$$
\LL_\xi V^M=
\xi^ND_NV^M+Z^{MN}{}_{PQ}D_N\xi^PV^Q\komma\Eqn\CovariantLDef
$$
where $D=\*+\Gamma$. One only has to consider the inhomogeneous
transformation of a connection $\Gamma$. 
We use the convention $D_MV_N=\*_MV_N+\Gamma_{MN}{}^PV_P$.
Denote any inhomogeneous
transformation (deviation from tensorial) by
$\Delta_\xi \phi\equiv\delta_\xi\phi-\LLO_\xi\phi$. Then
$$
\Delta_\xi\Gamma_{MN}{}^P=Z^{PQ}{}_{RN}\*_M\*_Q\xi^R\punkt\eqn
$$
Inserting this in the transformation (\CovariantLDef) leads to
$$
\Delta_\eta(\LL_\xi V^M)=
-\left(Z^{MN}{}_{TQ}Z^{TP}{}_{RS}+Z^{MP}{}_{RQ}\d^N_S\right)
\*_N\*_P\eta^R\xi^SV^Q\punkt\eqn
$$
If this vanishes, with the help of the section condition, the
transformation is covariant, and $\LL=\LLO$ for a torsion-free
connection. This can be shown explicitly for all the cases up to
$n=7$ (eq. (\NonLinTwo) above).
We should stress that the reasoning only holds for transformations
constructed with naked derivatives.


\section\EEightFailure{Beginning of a geometric construction for
$E_8$}Let us now reconsider the $E_8$ case.
The coordinate representation $R_1$ is the adjoint, and $R_2$ is 
${\bf1}\oplus{\bf3875}$, leaving only ${\bf27000}$ in the symmetrised
product of two derivatives. Any solution implies that also the
antisymmetrised ${\bf248}$ vanishes. (For more details about $E_8$
representations and tensor products, see \eg\
ref. [\KoepsellNicolaiSamtleben].) 
It is known that the natural candidate for a transformation,
$$
\LLO_\xi V^M=
\xi^N\*_NV^M+Z^{MN}{}_{PQ}\*_N\xi^PV^Q\komma\Eqn\EEightCandidateTransf
$$
with $Z_{MN}{}^{PQ}=-f_{AM}{}^Qf^A{}_N{}^P+\d_M^Q\d_N^P$, does not
lead to a closed algebra.
On the other hand, there is no ``better'' form with naked
derivatives. Eq. (\EEightCandidateTransf) has precisely the property
that it can be written in terms of a $Y$ tensor, projecting on modules
vanishing due to the
section condition:
$$
\LLO_\xi V^M=L_\xi V^M+(14P_{(\bf3875)}-30P_{(\bf248)}
         +62P_{(\bf1)})^{MN}{}_{PQ}\*_N\xi^PV^Q\punkt\eqn
$$
A direct calculation 
[\BermanCederwallKleinschmidtThompson,\HohmSamtleben] shows that 
$$
[\LLO_\xi,\LLO_\eta]V^M=\LLO_{{1\over2}(\LLO_\xi\eta-\LLO_\eta\xi)}V^M
+\fr2f^{MN}{}_Pf^Q{}_{RS}(\*_N\*_Q\xi^R\eta^S-\*_N\*_Q\eta^R\xi^S)V^P
\punkt\eqn
$$
The anomalous term takes the form of a local ${\frak e}_8$
transformation with a parameter carrying an index obeying the section
condition. This was used by Hohm and Samtleben in ref. [\HohmSamtleben]. 

In view of the connection between covariance and closure discussed
in section \CovarianceAndClosure, 
let us examine the failure in geometric terms. Here it is
important to keep in mind that the analysis is performed with respect
to the na\"\i ve transformation (\EEightCandidateTransf), which is
known to have problems. The considerations concerning covariance
etc. are not the final ones, only helpful steps on the way.

The occurrence of a two-derivative term points strongly to the
transformation of a connection [\Rosabal]. Let us perform a geometric check of
the covariance (and, thereby, the closure), which we know will fail,
but which will give interesting information. 
Define torsion as the part of the connection $\Gamma$ that transforms
covariantly under the transformation (\EEightCandidateTransf).
Since the connection is a one-form taking values in ${\frak
e}_8\oplus\RR$, the possible $E_8$ modules in $\Gamma_{MN}{}^P$ are
$$
{\bf248}\otimes({\bf1}\oplus{\bf248})=
{\bf248}\oplus({\bf1}\oplus{\bf3875}\oplus{\bf27000})_s
\oplus({\bf248}\oplus{\bf30380})_a\komma\Eqn\GammaModules
$$
where the subscripts denote the symmetric and antisymmetric tensor
products of the two ${\bf248}$'s.
In order for the connection to produce a covariant derivative, its
transformation must contain an inhomogeneous term
$$
\Delta_\xi\Gamma_{MN}{}^P\equiv(\delta_\xi-\LLO_\xi)\Gamma_{MN}{}^P
=Z^{PQ}{}_{RN}\*_M\*_Q\xi^R\punkt\eqn
$$
This expression of course transforms in
${\bf248}\otimes({\bf1}\oplus{\bf248})$, but thanks to the section
condition it must also lie in
${\bf27000}\otimes{\bf248}$. The irreducible modules in the overlap
are
${\bf248}\oplus{\bf27000}\oplus{\bf30380}$. This means that the remaining 
${\bf1}\oplus{\bf248}\oplus{\bf3875}$ will transform covariantly, and are
torsion. 
The torsion combination of the two ${\bf248}$'s turns out to be the
linear combination
$\Gamma_{NM}{}^N+\fr{248}\Gamma_{MN}{}^N$ of the two ${\bf248}$'s
appearing in eq. (\GammaModules).

What now goes wrong with the proposed na\"\i ve transformation
(\EEightCandidateTransf) is that even a torsion-free connection does
not drop out of the covariantised expression
$$
\LL^{(\Gamma)}_\xi V^M=\xi^ND_NV^M+Z^{MN}{}_{PQ}D_N\xi^PV^Q\komma\eqn
$$
as it did for $n\leq7$. Using the projection operators of the
appendix, a connection is torsion-free if 
$$
\eqalign{
&f^{MN}{}_P\Gamma_{MN}{}^P=0\komma\cr
&\bigl(f_{A(M}{}^Pf^A{}_{N)}{}^Q-2\delta_{(M}^P\delta_{N)}^Q\bigr)
        f_P{}^R{}_S\Gamma_{QR}{}^S=0\komma\cr
&\Gamma_{NM}{}^N+\fr{248}\Gamma_{MN}{}^N=0\cr
}
\Eqn\TorsionFreeConnection
$$
(the last relation is a linear combination of the two ${\bf248}$'s).
The transformation fails to be covariant, and in light of the previous
section, the algebra will not close.
We can investigate precisely how the different irreducible modules in
$\Gamma$ enter in $\LL^{(\Gamma)}_\xi V^M$. For this purpose we use
the projection operators listed in the appendix.
It now turns out that a torsion-free connection, \ie, one satisfying
eq. (\TorsionFreeConnection), with vanishing
${\bf1}\oplus{\bf248}\oplus{\bf3875}$, but remaining components in
${\bf248}\oplus{\bf27000}\oplus{\bf30380}$, will satisfy
$$
\LL^{(\Gamma)}_\xi V^M=\LLO_\xi V^M
-\fr{60}f^{MQ}{}_Pf_N{}^R{}_S\Gamma_{QR}{}^S\xi^NV^P\Eqn\LLGammaLLO
$$
(the number 60 is twice the Coxeter number).
The inhomogeneous transformation of a connection gives precisely the
failure of closure.
The extra term in eq. (\LLGammaLLO) 
is an ${\frak e}_8$ transformation of $V$ with parameter 
$$
\Sigma_{\xi M}=-\fr{60}f^N{}_{PQ}\Gamma_{MN}{}^P\xi^Q\punkt\Eqn\SigmaDef
$$
Instead of the equality of $\LL^{(\Gamma)}_\xi$ with $\LLO_\xi$, that
holds for $n\leq7$, we now have
$$
\LL^{(\Gamma)}_\xi=\LLO_\xi+\ad\Sigma_\xi\punkt\Eqn\LPlusSigma
$$
Note that eq. (\LPlusSigma) holds for a torsion-free connection. But
since torsion, per definition, is covariant, not only
$\LL^{(\Gamma)}_\xi$, but also the right hand side of
eq. (\LPlusSigma), is covariant, with $\Sigma$ constructed from {\it
any} connection as in eq. (\SigmaDef). 

So far, the ``geometric'' considerations have been performed with
respect to the transformations $\LLO_\xi$.
We know that $\LL^{(\Gamma)}_\xi$, per definition, is covariant
with respect to $\LLO_\xi$, but this is not the goal (and it is not
really a statement that makes geometric sense). We need an
expression for the transformations 
that is covariant with respect to {\it itself} (like for $n\leq7$).
Can $\LL^{(\Gamma)}_\xi$ have this property?

We drop the superscript ``$(\Gamma)$'', and let 
$$
\LL_\xi=\LLO_\xi+\ad\Sigma_\xi\punkt\Eqn\LLDef
$$ 
This is our candidate transformation for $n=8$. It is highly
unconventional in that it depends on a connection.


\section\EEightCovariance{Covariance and algebra}Before checking for
the covariance of the transformation (\LLDef) with 
respect to itself, we would like to consider connections in this
setting. All connections above are connections transforming with the
appropriate inhomogeneous terms under $\LLO$, not $\LL$. We want to
check how a covariant derivative $D=\*+\Gamma$ must transform in order
to take tensors to tensors. 
By a tensor we mean an object that transforms under scaling as is
induced by the transformation of a vector (whose scaling weight we
normalise to 1). Tensor densities may transform with other weights.
The connection $\Gamma$ is not necessarily
the same one as is used in $\Sigma$. It is straight-forward to check that
the presence of the $\Sigma$ term in the transformation leads to one
more inhomogeneous term in the transformation of a connection:
$$
\Delta_\xi\Gamma_{MN}{}^P
\equiv(\delta_\xi-\LL_\xi)\Gamma_{MN}{}^P
=Z^{PQ}{}_{RN}\*_M\*_Q\xi^R
          +f_N{}^{PQ}\*_M\Sigma_{\xi Q}\punkt\Eqn\TorsionTransform
$$
As mentioned, this transformation rule holds for any connection, in
particular for the one used to define $\Sigma$. This can be used quite
trivially to
obtain the inhomogeneous transformation of $\Sigma$ on the form
$$
\Delta_\xi\Sigma_\eta=X^{\xi,\eta}+Y^{\xi,\eta}\komma\Eqn\SigmaInhom
$$
where the inhomogeneous terms $X$ and $Y$,
$$
\eqalign{
X^{\xi,\eta}_M&=f^N{}_{PQ}\*_M\*_N\xi^P\eta^Q\komma\cr
Y^{\xi,\eta}_M&=\*_M\Sigma_{\xi N}\eta^N\komma\cr
}\eqn
$$
have been introduced for convenience in the following calculation.


When now the (candidate) transformation (\LLDef) is no longer linear, but
contains explicit fields (connection) through $\Sigma$, closure and
covariance are not equivalent. Covariance is essential for the
geometric framework, so we will first focus on that, and then check
what the implications for the algebra are.
The condition for
covariance of this expression with respect to the transformations it
generates reads
$$
\hat\delta_\eta(\LL_\xi V)
=\delta_\eta(\LL_\xi V)+\LL_{\LL_\eta\xi}V=\LL_\eta\LL_\xi V\punkt
\Eqn\CovCondZero
$$
We will go through the full check of covariance, even if part of it
(the covariance of $\LL$ with respect to $\LLO$)
follows from the considerations above.
Remember that $\delta$ only acts on fields. The second term on the left
hand side is the additional covariant transformation of the
parameter. 
The $\LL$'s, on the other hand, are just
operators, acting on everything on the right.
This can be rewritten as
$$
([\LL_\xi,\LL_\eta]+\ad(\delta_\eta\Sigma_\xi)+\LL_{\LL_\eta\xi})V=0
\punkt\Eqn\CovCondOne
$$
We use only the Leibniz rule (\ie, ``the product rule'') for $\LLO$ and the
Jacobi identity for the adjoint action, which together provide the Leibniz
rule for $\LL$. 
After inserting the split (\LLDef) into
eq. (\CovCondOne) and throwing away some cancelling terms, we get the
condition
$$
\eqalign{
0&=[\LLO_\xi,\LLO_\eta]+\LLO_{\LLO_\eta\xi}\cr
&\quad+\ad(\delta_\eta\Sigma_\xi+\Sigma_{\LLO_\eta\xi+[\Sigma_\eta,\xi]}
    -\LLO_\eta\Sigma_\xi-[\Sigma_\eta,\Sigma_\xi])\cr
&\quad+\LLO_{[\Sigma_\eta,\xi]}+\ad(\LLO_\xi\Sigma_\eta)\punkt\cr
}\Eqn\CovCondTwo
$$

The first line, with the parameters in this order, and no
(anti-)symmetrisation understood, states the failure of covariance for
$\LLO$ (with respect to itself). Let us check it first (although it
follows from the calculation above). 
It is convenient to introduce 
$$
\eqalign{
\leftbr\xi,\eta\rightbr^\circ&=\fr2(\LLO_\xi\eta-\LLO_\eta\xi)\komma\cr
\leftpar\xi,\eta\rightpar^\circ&=\fr2(\LLO_\xi\eta+\LLO_\eta\xi)\punkt\cr
}\eqn
$$
The action of $\LLO_{\leftpar\xi,\eta\rightpar^\circ}$ on a
vector does not vanish. $\leftpar\xi,\eta\rightpar^{\circ M}$
contains, in addition to reducibility in ${\bf1}\oplus{\bf3875}$,
a part 
$$
\fr4f_A{}^{MN}f^A{}_{PQ}(\*_N\xi^P\eta^Q+\*_N\eta^P\xi^Q)\komma\eqn
$$
which leads to 
$$
\LLO_{\leftpar\xi,\eta\rightpar^\circ}
=-\fr2\ad(X^{\xi,\eta}+X^{\eta,\xi})\punkt
\eqn
$$
We know from earlier that
$$
[\LLO_\xi,\LLO_\eta]=\LLO_{\leftbr\xi,\eta\rightbr^\circ}
+\fr2\ad(X^{\xi,\eta}-X^{\eta,\xi})\komma\eqn
$$
so the full first row or eq. (\CovCondTwo) becomes
$$
[\LLO_\xi,\LLO_\eta]+\LLO_{\LLO_\eta\xi}
=[\LLO_\xi,\LLO_\eta]-\LLO_{\leftbr\xi,\eta\rightbr^\circ}
         +\LLO_{\leftpar\xi,\eta\rightpar^\circ}
=-\ad X^{\eta,\xi}\punkt\Eqn\AnomalyOne
$$
It is essential that the $\*^2\xi\eta$ terms 
from $X^{\xi,\eta}$ cancel, as they can not appear in the additional
inhomogeneous terms coming from the transformation of a connection.

The second line of eq. (\CovCondTwo) states the deviation of
$\Sigma_\xi$ alone from 
being covariant under the full transformation $\LL_\eta$ (note that
the weight of $\Sigma_\xi$ is 0, since it is constructed 
from a vector $\xi$ with weight 1 and a connection with weight $-1$).
We have already calculated this in eq. (\SigmaInhom); the second line becomes
$\ad(X^{\eta,\xi}+Y^{\eta,\xi})$. We note that the $X$ terms cancel
between the first and second lines. This is not surprising, and only a
consequence of the covariance of $\LL$ under $\LLO$ discussed above.

Finally, the cross terms of the third line. 
Its first term contains a translation term,
which acting on a vector $V_M$ gives
$f^{NP}{}_Q\Sigma_{\eta P}\xi^Q\*_NV_M$.
It must disappear if cancellation with the remaining terms, which are
${\frak e}_8$ transformations, is to be possible.
If $\Sigma_M$ fulfills the section condition, the translation term
goes away. This means that the connection used to define $\Sigma$ has
to respect the section condition regarding its first index. Unless
such an extra condition is introduced by hand 
(which may be possible)\foot{Such a condition is used \eg\ in
ref. [\GodazgarGodazgarNicolai]. 
It seems to be allowed, since the concept of torsion can
be extended to allow for such a choice to be made, in view of the
transformation (\TorsionTransform). We do not find it
practical, however, since its solution demands splitting the
connection into $GL(n)$ modules.},
there is only one possibility, namely the Weitzenb\"ock connection
$$
W_{MN}{}^P=-(\*_MEE^{-1})_N{}^P\komma\eqn
$$
defined for a generalised vielbein $E$.
It is a flat but torsionful connection. From now on, we will assume
that $\Sigma$ is constructed with the Weitzenb\"ock connection,
$$
\Sigma_{\xi M}=-\fr{60}f^N{}_{PQ}W_{MN}{}^P\xi^Q
=\fr{60}f^N{}_{PQ}(\*_MEE^{-1})_N{}^P\xi^Q\punkt\Eqn\SigmaDef
$$ 
With this assumption, one easily derives the identity
$$
\LLO_{[\Sigma_\eta,\xi]}+\ad(\LLO_\xi\Sigma_\eta)
=-\ad Y^{\eta,\xi}\punkt\eqn
$$

All anomalous terms thus cancel, 
and $\LL$ is covariant with
respect to itself, which is quite remarkable. What
was expected, and more or less trivial, was that $\LL$ should be
covariant with respect to $\LLO$. This happens for any connection, not just
Weitzenb\"ock. That the other anomalous terms (with $Y$) also cancel is more
surprising. 

It should be noted that even if we needed the Weitzenb\"ock connection
in the definition of the transformation, any connection, for example a
torsion-free connection compatible with the covariant constancy of a
vielbein, may be used for the construction of covariant derivatives.
Which part of the connection is now torsion, in the sense that it
transforms covariantly under the new transformations $\LL_\xi$?
With the previous definition, under $\LLO_\xi$, we had torsion in 
${\bf1}\oplus{\bf248}\oplus{\bf3875}$. 
A direct inspection of the second inhomogeneous term in
eq. (\TorsionTransform) shows that precisely these modules drop out
due to the section condition, since $\Sigma$ is formed from the
Weitzenb\"ock connection. The notion of torsion remains
unchanged. This is of course essential when it comes to defining
torsion-free connections, curvatures etc.


We can now consider the commutator of two transformations. We get
$$
\delta_\eta(\delta_\xi V)=\delta_\eta(\LL_\xi V)
=\LL_\xi\LL_\eta V+\ad(\delta_\eta\Sigma_\xi)V\punkt\eqn
$$
A simple comparison with eq. (\CovCondOne) gives at hand that
$$
\delta_\eta(\delta_\xi V)-\delta_\xi(\delta_\eta V)
=(\LL_{\leftbr\xi,\eta\rightbr}-\ad(\delta_{[\xi}\Sigma_{\eta]}))V
\punkt\eqn
$$
The expression 
$$
\delta_{[\xi}\Sigma_{\eta]}
=\LL_{[\xi}\Sigma_{\eta]}-\Sigma_{\leftbr\xi,\eta\rightbr}
+(X+Y)^{[\xi,\eta]}\eqn
$$ 
should now be a tensor.
Note that $\LL_\xi\Sigma_\eta=\LLO_\xi\Sigma_\eta$, but that the
second term in 
$\Sigma_{\leftbr\xi,\eta\rightbr}=\Sigma_{\leftbr\xi,\eta\rightbr^\circ}
+\Sigma_{{1\over2}([\Sigma_\xi,\eta]-[\Sigma_\eta,\xi])}$ can not be
dropped, and will be quadratic in connections.

The tensorial property seems intuitively natural, considering that it
is a variation of a connection. It follows directly from the
definition of the transformation of a connection,
$$
\delta_\xi\Gamma_{MN}{}^P=\LL_\xi(D_MV_N)-D_M\LL_\xi V_N\komma\eqn
$$
but it may of course be spelt out more concretely. The result of
ref. [\HohmSamtleben] that the commutator of two generalised
diffeomorphisms contains a section-restricted ${\frak e}_8$
transformation remains, but has been given a covariant formulation. In
the present formalism, there is however no need to introduce a
separate connection for the ${\frak e}_8$
transformations, since the geometric connection already transforms in
the appropriate way.


\section\TorsionSpinConnCurvature{Torsion, spin connection and
curvature}Let us now consider curvature, and try to construct it for a
general connection. This attempt, which will not be entirely
successful, for reasons we will come back to, goes along the same
lines as constructions for lower $n$ in several papers, \eg\ refs. 
[\CoimbraStricklandWaldram,\CoimbraStricklandWaldramII,\AldazabalGranaMarquesRosabal,\CederwallEdlundKarlsson]
The starting point would be
the inhomogeneous transformation of the connection,
$$
\Delta_\xi\Gamma_{MN}{}^P=Z^{PQ}{}_{RN}\*_M\*_Q\xi^R
+f_N{}^{PQ}\*_M\Sigma_{\xi Q}\punkt\eqn
$$
It leads to the transformation of the derivative of a connection:
$$
\eqalign{
\Delta_\xi\*_M\Gamma_{NP}{}^Q&=Z^{QR}{}_{SP}\*_M\*_N\*_R\xi^S
+f_P{}^{QR}\*_M\*_N\Sigma_{\xi R}\cr
&+\Delta_\xi\Gamma_{MR}{}^Q\Gamma_{NP}{}^R
-\Delta_\xi\Gamma_{MN}{}^R\Gamma_{RP}{}^Q
-\Delta_\xi\Gamma_{MP}{}^R\Gamma_{NR}{}^Q\punkt\cr
}\eqn
$$
The first two terms can be removed by antisymmetrisation $[MN]$,
leading to 
$$
\Delta_\xi\left(2\*_{[M}\Gamma_{N]}+2\Gamma_{[M}\Gamma_{N]}\right)_P{}^Q
=-2\Delta_\xi\Gamma_{[MN]}{}^R\Gamma_{RP}{}^Q\punkt\Eqn\CurvInterm
$$
In ordinary geometry, the right hand side vanishes, since
$\Gamma_{[MN]}{}^P$ is torsion, and the covariance of the ordinary
Riemann tensor is obtained. Here we need to use the identity for
torsion, eq. (\LLGammaLLO), which states that
$$
T_{MN}{}^P\equiv\Gamma_{MN}{}^P+Z^{PQ}{}_{RN}\Gamma_{QM}{}^R
-\fr{60}f^{PQ}{}_Nf_M{}^R{}_S\Gamma_{QR}{}^S\Eqn\TorsionEq
$$
is torsion. It can equivalently be written
$$
T_{MN}{}^P=2\Gamma_{[MN]}{}^P+Y^{PQ}{}_{RN}\Gamma_{QM}{}^R
-\fr{60}f^{PQ}{}_Nf_M{}^R{}_S\Gamma_{QR}{}^S\punkt\eqn
$$
Using $\Delta_\xi T=0$ in the right hand side of eq. (\CurvInterm)
gives
$$
\eqalign{
&\Delta_\xi\left(2\*_{[M}\Gamma_{N]}+2\Gamma_{[M}\Gamma_{N]}\right)_P{}^Q\cr
&\qquad=Y_{NR}{}^{ST}\Delta_\xi\Gamma_{TM}{}^R\Gamma_{SP}{}^Q
-\fr{60}f_N{}^{RS}f_M{}^T{}_U\Delta_\xi\Gamma_{ST}{}^U\Gamma_{RP}{}^Q\punkt\cr
}\Eqn\CurvFirstStep
$$
The first term on the right hand side can be written as a total
transformation if contracted with $\delta_Q^N$ and symmetrised
$(MP)$. 
This reflects the
usual phenomenon that a full ``Riemann'' 4-index tensor can not be
formed, only a 2-index ``Ricci'' tensor. We then have
$$
\eqalign{
&\Delta_\xi\left.\left(2\*_{[M}\Gamma_{N]}
          +2\Gamma_{[M}\Gamma_{N]}\right)_P{}^N\right\vert_{(MP)}\cr
&\qquad=Y_{NR}{}^{ST}\Delta_\xi\Gamma_{T(M}{}^R\Gamma_{|S|P)}{}^N
-\fr{60}f_N{}^{RS}f^T{}_{U(M}\Gamma_{|R|P)}{}^N\Delta_\xi\Gamma_{ST}{}^U\punkt\cr
}\eqn
$$
The first term on the right hand side can be written as
$\fr2Y_{NR}{}^{ST}\Delta_\xi(\Gamma_{T(M}{}^R\Gamma_{|S|P)}{}^N)$. The
second term does not display this symmetry, but we note that it can be
expressed as 
$-\fr{60}f_N{}^{RS}f^T{}_{U(M}\Delta_\xi(\Gamma_{|R|P)}{}^NW_{ST}{}^U)$,
since $\Delta_\xi W=\Delta_\xi\Gamma$, and the inhomogeneous
transformation of the first $\Gamma$ drops out due to $W$ (and not
only its inhomogeneous variation) satisfying the section condition.
We have thus showed that the ``Ricci'' tensor
$$
\eqalign{
R_{MN}&=\*_{(M}\Gamma_{|P|N)}{}^P-\*_P\Gamma_{(MN)}{}^P
+\Gamma_{(MN)}{}^Q\Gamma_{PQ}{}^P-\Gamma_{P(M}{}^Q\Gamma_{N)Q}{}^P\cr
&-\fr2Y^{PQ}{}_{RS}\Gamma_{PM}{}^R\Gamma_{QN}{}^S
+\fr{60}f_P{}^{QR}f^S{}_{T(M}\Gamma_{|Q|N)}{}^PW_{RS}{}^T\cr
}\Eqn\FakeRicci
$$
transforms covariantly. It vanishes for the Weitzenb\"ock connection,
as expected. The appearance of $W$ in the expression for the curvature
is peculiar, but maybe not more so than its appearance in the
transformations. The value of the expression (\FakeRicci) is however
doubtful --- it is quite likely that it will lead to curvature in the
$Spin(16)/\ZZ_2$ representation ${\bf128}$, the candidate for the
equation of motion for the generalised metric or vielbein, which
contains undefined connection. We have however not been able to
strictly show that this is the case, so it is still an open question
whether the tensor of eq. (\FakeRicci) can give rise to \eg\ a
well-defined scalar curvature.
We also note that this kind of ``fake curvature'' with explicit $W$'s,
can be constructed also for a full Riemann (4-index) tensor, by using
the section condition on the right hand side of eq. (\CurvFirstStep).


It is our impression that the last term in the torsion, which is
traced back to the ``mismatch'' between the transformations and the
torsion, causing the failure of closure of the transformation with
naked derivatives ($\LLO$), cannot be compensated for by a pure
``$\Gamma\Gamma$'' term. The absence of some (at least 2-index)
curvature for an arbitrary connection may not be a disaster,
however. The important thing in the end is to have some curvature in
${\bf128}$ of $Spin(16)/\ZZ_2$, which contains only connection which
is well defined by compatibility, and this is far from
excluded. Compatible and solvable (affine or spin) connections will contain
explicit derivatives, which means that the section condition can be at
work, to a higher degree than above, when constructing curvature.

We therefore change our strategy, and focus on
solving, as far as possible, for a spin connection
$\Omega_{MA}{}^B$. 
The covariant constancy of the generalised vielbein reads
$$
D_ME_N{}^A=\*_ME_N{}^A+\Gamma_{MN}{}^PE_P{}^A-E_N{}^B\Omega_{MB}{}^A=0\punkt
\eqn
$$
Using the vanishing torsion in $\Gamma$ the equation for the spin
connection becomes
$$
T(E\Omega E^{-1}+W)=0\komma\eqn
$$
where $T(\Gamma)$ is the torsion combination of eq. (\TorsionEq). This
equation contains the $E_8$ modules with $Spin(16)$ decomposition:
$$
\eqalign{
{\bf1}&\rightarrow{\bf1}\cr
{\bf248}&\rightarrow{\bf120}\oplus{\bf128}\cr
{\bf3875}&\rightarrow{\bf135}\oplus{\bf1820}\oplus{\bf1920}\cr
}\eqn
$$
The modules appearing in the decomposition of ${\bf3875}$ are a
symmetric traceless tensor, a 4-form, and
a $\Gamma$-traceless vector-cospinor.
The modules that can appear in $\Omega$ are (with flattened form
index):
$$
\EqMatrix{
\Omega_{ab}\hfill&:&{\bf120}\otimes{\bf120}
        &=&{\bf1}\oplus{\bf120}\oplus{\bf135}\oplus{\bf1820}
\oplus\Gray{\bf5304}\oplus\Gray{\bf7020}\hfill\cr
\Omega_{\alpha}\hfill&:&{\bf128}\otimes{\bf120}
        &=&{\bf128}\oplus{\bf1920}\oplus\Gray{\bf13312}\hfill
}\eqn
$$
The modules in {\bf black} may be solved for, and the ones in 
\Gray{{\bf grey}}, which in
order of appearance are 
traceless tensors of types $\yng(2,2)$ and $\yng(2,1,1)$ and a
$\Gamma$-traceless 2-form-spinor, remain undetermined. 
In any equation used for fields or transformations, one should in the
end only use covariant derivatives where the undefined
connection components drop out. 
Examples of such well-defined covariant derivatives are
${\bf16}\rightarrow{\bf16}$,
${\bf128'}\rightarrow{\bf128'}$ and
${\bf16}\leftrightarrow{\bf128'}$. This will be important when
considering local supersymmetry (which is beyond the scope of this paper).


In order to solve for the solvable part of the spin connection, we
need to decompose our tensors into $Spin(16)$ modules. A choice for
the $E_{8(8)}$ structure constants and metric corresponding to the normalisation
used is
$$
\eqalign{
&f^{ab,cd}{}_{ef}=-2\sqrt2\delta^{[a}_{[e}\delta^{b][c}\delta^{d]}_{f]}\komma\cr
&f^{ab,\alpha}{}_\beta=\fr{2\sqrt2}(\Gamma^{ab})^\alpha{}_\beta\komma\cr
&\eta_{ab,cd}=-\delta^{ab}_{cd}\komma\quad
\eta_{\alpha\beta}=\delta_{\alpha\beta}\cr
}\Eqn\StructureConstants
$$
(the overall normalisation is $\sqrt2$ times the one most commonly
used in the physics literature).
It will be useful to know the form of the decomposition of the
projections of two-index tensors on ${\bf248}$ (antisymmetric) and
${\bf1}\oplus{\bf3875}$ used in the section condition, and also in the
torsion. An antisymmetric tensor $A_{AB}$ will have vanishing
component in ${\bf248}\rightarrow{\bf120}\oplus{\bf128}$ if
$$
\eqalign{
&A_{ac,bc}-\fr8(\Gamma_{ab})^{\alpha\beta}A_{\alpha\beta}=0\komma\cr
&(\Gamma^{ab})_\alpha{}^\beta A_{ab,\beta}=0\komma\cr
}\Eqn\AEquation
$$
and the part of a symmetric tensor $S_{AB}$ in ${\bf1}\oplus{\bf3875}
\rightarrow{\bf1}\oplus{\bf135}\oplus{\bf1820}\oplus{\bf1920}$
vanishes if
$$
\eqalign{
&S_{ai,bi}-\fr{16}\delta_{ab}S_{\alpha\alpha}=0\komma\cr
&S_{[ab,cd]}+\fr{48}(\Gamma_{abcd})^{\alpha\beta}S_{\alpha\beta}=0\komma\cr
&(\Gamma^i)_{\dot\alpha}{}^\alpha S_{ai,\alpha}
  -\fr{16}(\Gamma_a\Gamma^{ij})_{\dot\alpha}{}^\alpha S_{ij,\alpha}=0\punkt\cr
}\Eqn\SEquation
$$
As a consistency check of our structure constants
(\StructureConstants), the middle equation turns up both in the
$ab,cd$ and the $\alpha\beta$ part of the projection on ${\bf3875}$,
with the same combination, and the $\Gamma$-tracelessness of the last
equation is reproduced correctly.

We can now solve for the well-defined part of the spin connection. If
we use a one-index notation both for the spin connection and the $E_8$
part of the
Weitzenb\"ock connection, so that (with flattened indices)
$$
\eqalign{
W_{AB}{}^C&=f_B{}^{CD}W_{AD}+\delta_B^Cw_A\komma\cr
\Omega_{AB}{}^C&=f_B{}^{C,ab}\Omega_{A,ab}\komma\cr
}\eqn
$$
the solution is obtained using eqs. (\TorsionFreeConnection), (\AEquation) and (\SEquation), and reads:
$$
\EqMatrix{
\hfill{\bf1}\oplus{\bf135}\oplus{\bf120}:
&\hfill\Omega_{ai,bi}
&=&-W_{ai,bi}+\fr8(\Gamma_{ab})^{\alpha\beta}W_{\alpha\beta}
        +\fr{16}\delta_{ab}W_{\alpha\alpha}-\fr{\sqrt2}w_{ab}\komma
                        \hfill\cr
\hfill{\bf1820}:&\hfill\Omega_{[ab,cd]}
&=&-W_{[ab,cd]}
        -\fr{48}(\Gamma_{abcd})^{\alpha\beta}W_{\alpha\beta}\komma
                        \hfill\cr
\hfill{\bf128}\oplus{\bf1920}:
&\hfill(\Gamma^i)_{\dot\alpha}{}^\alpha\Omega_{\alpha,ai}
&=&-(\Gamma^i)_{\dot\alpha}{}^\alpha(W_{\alpha,ai}+W_{ai,\alpha})
        +\fr8(\Gamma_a\Gamma^{ij})_{\dot\alpha}{}^\alpha W_{ij,\alpha}
                        \hfill\cr
&&& -\fr{2\sqrt2}(\Gamma_a w)_{\dot\alpha}\punkt\hfill
}\Eqn\OmegaSolution
$$
The right hand sides represent the torsion components $-T_{a,b}$,
$-T_{abcd}$ and $-T_{a\dot\alpha}$ of the Weitzenb\"ock connection,
sometimes referred to as fluxes (not to be confused with the
torsion of the affine connection which we have chosen to vanish).

Since $\Omega$ is constructed from derivatives of the connection,
there will be implications from the section condition that may help in
the construction of curvature. We have checked, through a rather long
calculation, that there is a
spinorial constraint
$$
\fr{24}(\G^{abcd})_\alpha{}^\beta\Omega_{ab,cd}\*_\beta
+\fr2(\Gamma^{ab})_\alpha{}^\beta\Omega_{ai,bi}\*_\beta
-\Fr5{12}\Omega_{ij,ij}\*_\alpha
-\fr{12}(\Gamma^c\Gamma^{ab}\Gamma^d)_\alpha{}^\beta\Omega_{\beta,cd}\*_{ab}
=0\punkt\eqn
$$
This is shown by direct insertion of the solution (\OmegaSolution) for $\Omega$
into an Ansatz, and using the section condition between the derivative
and the first index on $W$ (or the index on $w$) on the forms
(\AEquation,\SEquation). The relation relies on the Fierz identity
$$
F^{\alpha\beta\gamma\delta}=F^{\alpha\delta\gamma\beta}\komma\eqn
$$
where
$$
F^{\alpha\beta\gamma\delta}=
\fr{24}(\Gamma^{abcd})^{\alpha\beta}(\Gamma_{abcd})^{\gamma\delta}
-3(\Gamma^{ab})^{\alpha\beta}(\Gamma_{ab})^{\gamma\delta}
+20\delta^{\alpha\beta}\delta^{\gamma\delta}\eqn
$$
(note the absence of $\Gamma^{(6)}$ and $\Gamma^{(8)}$, which
would lead to terms where the section condition can not be used).

Instead of attempting to construct curvature from $\Omega$, which
is possible (see below), we will take another approach, namely to
construct a scalar $K$ which is quadratic in the 
torsion of the Weitzenb\"ock connection. 
This scalar will play the r\^ole analogous to that of a curvature
scalar in an action. 
This procedure is possible in Einstein gravity, and has been used
earlier in extended geometry [\BermanBlairMalekPerry,\BlairMalek].
In addition to the restriction to torsion,
dictated by covariance, it is also important that the only parts of $W$
that appear are $W_{M\alpha}$ and $w_M$. Then the expression will be
invariant under local $Spin(16)$ transformations; when a variation of the
vielbein is performed, $\delta K$ will not contain 
$(E^{-1}\delta E)_{ab}$, only $(E^{-1}\delta E)_\alpha$ and
$\tr(E^{-1}\delta E)$.

Na\"\i vely, the first terms on the right hand sides of
eq. (\OmegaSolution) seem to present obstructions to such a
construction, but the section condition may (and will) help. 
The contribution of
these first terms must vanish altogether, both the quadratic and
linear ones. By considering the quadratic part, we first find
that only 2 out of the 4 possible scalars from 
$({\bf1}\oplus{\bf135}\oplus{\bf120}\oplus{\bf1820})^2$ are possible,
if the first indices are to arrange in ways that can give a
cancellation with $({\bf128}\oplus{\bf1920})^2$ using the section
condition. These combinations are
$$
\eqalign{
&W_{[ab,cd]}W_{ab,cd}-\Fr23W_{ai,bi}W_{bj,aj}+\fr6W_{ij,ij}W_{kl,kl}\cr
&\qquad=W_{[ab,|ab|}W_{cd],cd}+\fr6W_{ab,cd}W_{ab,cd}
        -\Fr23W_{ac,bd}W_{ab,cd}\cr
&\hbox{and}\quad W_{ai,bi}W_{aj,bj}\punkt\cr
}\eqn
$$
When they are matched to the two possible contributions from 
$({\bf128}\oplus{\bf1920})^2$, we obtain a unique combination, modulo
an overall constant, where
the terms quadratic in $W_{M,ab}$ cancel, namely
$$
\eqalign{
K&=T_{abcd}T_{abcd}+\Fr43T_{a,b}T_{a,b}-\Fr23T_{a,b}T_{b,a}+\fr6T_{a,a}T_{b,b}\cr
&-\Fr3{16}T_{a\dot\alpha}T_{a\dot\alpha}
-\fr{48}(\Gamma^{ab})^{\dot\alpha\dot\beta}T_{a\dot\alpha}T_{b\dot\beta}\punkt\cr
}\Eqn\CovariantScalar
$$
Then one has to check for the terms linear in $W_{M,ab}$. It turns out
that these cancel, through what looks like a long series of numerical
coincidences, using the section condition to switch the first indices
on a pair of $W$'s.

This shows that the covariant scalar (\CovariantScalar) can be written
in a (non-covariant) form, where each $T$ is replaced by a 
(non-tensorial) $\tilde T$
obtained by omitting the first terms from eq. (\OmegaSolution), \ie,
$$
\eqalign{
\tilde T_{a,b}&=-\fr8(\Gamma_{ab})^{\alpha\beta}W_{\alpha\beta}
        -\fr{16}\delta_{ab}W_{\alpha\alpha}+\fr{\sqrt2}w_{ab}\komma\cr
\tilde T_{abcd}&=\fr{48}(\Gamma_{abcd})^{\alpha\beta}W_{\alpha\beta}\komma\cr
\tilde T_{a\dot\alpha}&=(\Gamma^i)_{\dot\alpha}{}^\alpha W_{ai,\alpha}
        -\fr8(\Gamma_a\Gamma^{ij})_{\dot\alpha}{}^\alpha W_{ij,\alpha}
        +\fr{2\sqrt2}(\Gamma_a w)_{\dot\alpha}\punkt\cr
}\Eqn\TTilde
$$
If the $\tilde T$'s are used, covariance is not manifest. 
If the $T$'s are used, covariance is manifest, but local
$Spin(16)$ only arises thanks to the section condition. 
This behaviour seems to support our speculation that a curvature
tensor (in terms of only a general torsion-free connection $\Gamma$,
or in terms of a spin connection $\Omega$) does not exist before the
section condition is used on the solution of the compatibility
equation.
Since the scalar we have found is unique, it must coincide with the
one given in ref. [\HohmSamtleben].

A ``Ricci tensor'' in ${\bf1}\oplus{\bf128}$, 
governing the equation of motion for
the vielbein, is obtained by the formal variation of an ``action''
$$
S\sim\int|E|^{-{1\over248}}K\punkt\eqn
$$
The power of the determinant of the vielbein is dictated by the correct
weight of the integrand, allowing for partial integration
[\CederwallEdlundKarlsson]. 
In view of the $Spin(16)$-invariance, the variation will only contain
$(E^{-1}\delta E)_\alpha$ and $\tr(E^{-1}\delta E)$, 
and an ``Einstein tensor'' is obtained after
partial integration. It can in turn be contracted to
a scalar curvature $R$, such that $|E|^{-{1\over248}}K$ and
$|E|^{-{1\over248}}R$ differ by a total derivative.

\section\Reducibility{Covariant reducibility}As mentioned in the
introduction, the situation has been unclear not only concerning the
generalised diffeomorphisms for $E_8$, but also for a complete
understanding of the symmetries for lower $n$. Although the
generalised diffeomorphisms and the tensor formalism work fine, there
have been questions concerning the (infinite) reducibility, and its
associated (infinite) tower of ghosts for ghosts. This may seem like a
technical detail, but is important if an understanding of the global
properties is to be taken to the same level as the one for
$O(d,d)$. It concerns \eg\ the formulation of transition functions in terms
of gerbes [\BermanCederwallPerry]. As shown in
ref. [\CederwallEdlundKarlsson], there is a sequence of modules 
$$
R_1\leftarrow R_2\leftarrow\ldots\leftarrow R_{8-n}\komma\eqn
$$ 
where the arrows denote action of the derivative, for which
torsion-free connection drops out of the covariant derivatives. 
In this precise sense, the modules are analogous to forms in ordinary geometry.
These are modules appearing in tensor hierarchies (see \eg\
refs. [\deWitNicolaiSamtleben,\AldazabalGranaMarquesRosabalII]),  
coinciding with the
reducibility of the generalised diffeomorphisms 
[\BermanCederwallKleinschmidtThompson], with parameter in $R_1$, and
also of tensor fields, with parameters in higher $R_k$. The sequence
of $R_k$'s coincides with the generators of Borcherds superalgebras
[\PalmkvistHierarchy-\skipref\GreitzHowePalmkvist\skipref\PalmkvistTensor\skipref\HowePalmkvist-\CederwallPalmkvistBorcherds]. 
The sequence of ghosts, corresponding to reducibility, does not stop
where the connection-free window closes, however. It has been somewhat
disturbing that the complete ghost structure, formulated with naked
derivatives, has been non-covariant, beginning with 
$R_{8-n}\leftarrow R_{9-n}$, $R_{9-n}$ being the adjoint.
Of course, this is also an indication to why the problem comes all the way down
to the algebra of generalised diffeomorphisms for $n=8$.
In the light of the solution for $E_8$, the solution for lower $n$
becomes clear: The higher reducibilities should be formulated with
covariant derivatives containing the Weitzenb\"ock connection. This is
consistent, since such a covariant derivative automatically 
obeys the section condition.
We are not in a position to say what this implies for the gerbe
structure of exceptional geometry.

For the generalised diffeomorphisms of the present paper, the same
construction holds. Reducibility can be obtained covariantly, with
covariant derivatives containing the Weitzen\-b\"ock connection. At the
first step, $R_2={\bf1}\oplus{\bf3875}$, and a transformation
$\LL_\xi$ with $\xi^M=D_N\Lambda^{MN}$, $\Lambda\in R_2$, generates a
null transformation, $\LL_\xi V=0$.

\section\EhlersSymmetry{Ehlers symmetry}What has been done for $E_8$
above applies in spirit to enhanced symmetries arising on dimensional
reduction of gravity from $3+n$ to 3 dimensions due to the appearance of a dual
gravity field. This is the Ehlers symmetry
$SL(n+1)$. $E_{8(8)}$ of course contains an $SL(9)$ subgroup, but it
is possible to construct an extended geometry for all $n$. 
A series of models
built on $SL$ algebras was considered previously in
ref. [\ParkSuh]. It is different from the present one in the choice of
coordinate representation etc., and does not contain dual gravity.

The transformations are obtained exactly as the ones for $E_8$, with the
invariant $SL(n+1)$ tensor $Z$ given by the same expression, 
$Z_{MN}{}^{PQ}=-f_{AM}{}^Qf^A{}_N{}^P+\d_M^Q\d_N^P$. Tensors are most
conveniently written in fundamental indices, where the structure
constants are
$$
f_m{}^n{}_{,p}{}^q{}_{,r}{}^s=\delta^n_p\delta^q_r\delta^s_m
-\delta^n_r\delta^q_m\delta^s_p\komma\eqn
$$ 
and the invariant metric is
$$
\eta_m{}^n{}_{,p}{}^q=\delta^n_p\delta^q_m-\fr{n+1}\delta^n_m\delta^q_p
\punkt\eqn      
$$
The steps of sections \EEightFailure\ and \EEightCovariance\ can be
followed, where the numerical factor $\fr{60}$ is replaced by
$\fr{2(n+1)}$, in all cases equalling $\fr{2h}$, $h$ being the Coxeter
number. The projection operators can of course not be copied, but the
important relations used in the calculation can; it is straightforward
to show that the section condition also in this case implies
$$
\eqalign{
&\eta^{MN}\*_M\otimes\*_N=0\komma\cr
&f_M{}^{NP}\*_N\otimes\*_P=0\komma\cr
&(f_{AM}{}^Pf^A{}_N{}^Q-2\delta_{(M}^P\delta_{N)}^Q)\*_P\otimes\*_Q=0\punkt\cr
}\Eqn\EhlersSC
$$

The coordinate representation is also here the adjoint, with highest
weight Dynkin label $(10\ldots01)$. 
This choice can be inferred by considering the $SL(9)$ subgroup of
$E_{8(8)}$, but is ultimately motivated by the solution of the section
condition and 
of the field content in the coset described below. The representations
for coordinates and for the section condition also appears in the
classification of ref. [\StricklandConstable].

Although the symmetric product of two elements in the
adjoint generically contains four irreducible representations (instead
of three, for $E_8$), 
$$
\vee^2(10\ldots01)=(0\ldots0)\oplus(10\ldots01)
\oplus(010\ldots010)\oplus(20\ldots02)\komma\eqn
$$
the section condition removes all but the
largest one, the one
with highest weight twice the one of the adjoint. It effectively sets
to zero any contraction of fundamental indices, and a solution can be
taken as $\*_{m'}{}^0=\*_{m'}$ and all other derivatives vanishing, so
that fields locally depend on a set of coordinates $x^{m'}$,
$m'=1,\ldots,n$. The section condition reads
$$
\eqalign{
&\fr2(\*_m{}^n\otimes\*_p{}^q+\*_p{}^q\otimes\*_m{}^n)
=\*_{(m}{}^{(n}\otimes\*_{p)}{}^{q)}\cr
&\*_m{}^p\otimes\*_p{}^n=0\komma\cr
}\Eqn\EhlersSCIndex
$$ 
Eq. (\EhlersSC) is verified by the short calculation
$$
\eqalign{
&\fr2\bigl(\tr(A\*_1)\tr(B\*_2)+\tr(B\*_1)\tr(A\*_2)\bigr)\cr
&\qquad=\fr4\bigl(\tr(A\*_1)\tr(B\*_2)+\tr(B\*_1)\tr(A\*_2)
        +\tr(A\*_1B\*_2)+\tr(B\*_1A\*_2)\bigr)\,;\cr
&\tr([A,\*_1][B,\*_2])=\tr(A\*_1B\*_2+\*_1A\*_2B)
        =\tr(A\*_1)\tr(B\*_2)+\tr(B\*_1)\tr(A\*_2)\komma\cr
}\eqn
$$
using eq. (\EhlersSCIndex) to derive the last
equation in (\EhlersSC). 

Here, it is technically less complicated to isolate the dual gravity
field, which is the only field apart from the ordinary vielbein. An
$SL(n+1)/SO(n+1)$ vielbein can, fixing the $SO(n+1)$ gauge, be parametrised as
$$
E_m{}^a=\left[\matrix{|e|^{-1}&0\cr
                |e|^{-1}\phi_{m'}&e_{m'}{}^{a'}}\right]\komma\eqn
$$
where $\phi_{m'}$ represents the dual gravity field. The action on the
generalised vielbein by a restricted $SL(n+1)$ transformation
$$
T_m{}^n=\left[\matrix{0&0\cr
                t_{m'}&0}\right]\eqn
$$
amounts to a shift in $\phi$.
The dual gravity field does not carry any local degrees of
freedom, and it becomes clear that the restricted $SL(n+1)$
transformations should not be counted as removing any local degrees of
freedom beyond the ones removed by the generalised diffeomorphisms.

This is also verified by a counting of the effective number of degrees
of freedom removed by a generalised diffeomorphism. By the method of
ref. [\CederwallPalmkvistBorcherds] (see also
ref. [\BermanCederwallKleinschmidtThompson], where the corresponding
counting is performed for $E_8$ and for lower $n$), 
the relevant Borcherds algebra is
related to a bosonic object $\lambda$ in $(10\ldots01)$, constrained
so that the only module appearing at $\lambda^2$ is
$(20\ldots02)$. The partition function of $\lambda$ then becomes
$$
\eqalign{
Z_n(t)&=\sum\limits_{k=0}^\infty\hbox{dim}(k0\ldots0k)t^k
=\sum\limits_{k=0}^\infty\left({(n+k-1)!\over k!}\right)^2
                                              {n+2k\over n!(n-1)!}t^k\cr
&=(1-t)\,{}_2F_1(n+1,n+1;1;t)=(1-t)^{-2n}{}_2F_1(-n,-n;1;t)\cr
&=(1-t)^{-n}P_n\left({1+t\over1-t}\right)
=(1-t)^{-2n}\sum\limits_{i=0}^n{n \choose i}^{\!2}t^i\punkt\cr
}\eqn
$$
The partition functions for all $n$ are governed by a simple generating
function
$$
{\cal Z}(s,t)=\sum\limits_{n=0}^\infty Z_n(t)s^n
={1\over\sqrt{1-{2(1+t)s\over(1-t)^2}+{s^2\over(1-t)^2}}}\punkt\eqn
$$
$Z_n(t)$ is also the inverse partition function for the subalgebra of the
Borcherds superalgebra at positive levels
[\CederwallPalmkvistBorcherds], 
with generators in $R_k$. The effective number of gauge parameters,
modulo reducibility, equals the number of degrees of freedom in
$\lambda$, and may be read off as the power of the pole of the
partition function at $t=1$. The partition functions are 
rational functions with denominators
$(1-t)^{2n}$. Of the $2n$ gauge degrees of freedom, ordinary
diffeomorphisms and dual diffeomorphisms make up $n$ each, which
completely removes the local degrees of freedom for the dual gravity
field.

\section\Discussion{Discussion}We have shown how it is possible to
define covariant field-dependent generalised diffeomorphisms for
$E_8$, and used them to understand the dynamics in a geometric way. 
We would like to stress that the conclusions of Hohm and
Samtleben [\HohmSamtleben] remain true, but are given a geometric
framework. The solution also provides a covariant formulation of the
reducibility for lower $n$. A very similar construction, which we have
only sketched, is valid for the Ehlers symmetry $SL(n+1)$.

The transformations needed in order to achieve covariance and build a
tensor formalism are field-dependent, and depend on a generalised
vielbein through the 
Weitzenb\"ock connection $W_{MN}{}^P=-(\*_MEE^{-1})_N{}^P$.
This is very unconventional, but in this case necessary, and unlike
any previously encountered generalised diffeomorphisms.
Our analysis this far is entirely local, and it is not yet clear to us
what the consequences for global structures will be when such
transformations are used to relate overlapping patches. In double
geometry, a double manifold has a manifold structure before any fields
(\eg\ generalised vielbein or metric) are introduced, which on the
introduction of generalised metric data acquires a gerbe structure
(visible already at the level of the algebra)
[\BermanCederwallPerry]. In the present situation, no such distinction
is possible, since there is no way of constructing covariant
transformations without the presence of a vielbein. We would however
like to remind again that this is true also for the lower exceptional
cases. Although the field dependence there enters at higher ghost
levels, it will be necessary in order to understand the full
reducibility and the full ``gerbe''
structure. 

Hopefully, the present treatment can open the road towards higher $n$
and infinite-dimensional algebras, starting with $E_9$. 
There may be reason to wonder if the ``dual gravity barrier''
really has been broken, in a way that will persist for higher $n$, or
if new difficulties (apart from infinite dimensionality) will arise.
A reason for hope may be the unexpected covariance of the
field-dependent transformations, including the gauge symmetry for the
dual gravity field.
A reason for doubt, on the other hand, may be the observation that we
have not yet reached a situation where the dual gravity field becomes 
dynamical.  


\appendix{Projection operators for $E_8$ tensors}The tensor products
of two adjoint ${\bf248}$'s  
of $E_8$ contains the irreducible
modules ${\bf1}\oplus{\bf3875}\oplus{\bf27000}$ in the symmetric
part and ${\bf248}\oplus{\bf30380}$ in the antisymmetric
part. The projection operators on the irreducible modules are
$$
\eqalign{
&P_{({\bf1})}^{MN}{}_{PQ}=\fr{248}\eta^{MN}\eta_{PQ}\komma\cr
&P_{({\bf3875})}^{MN}{}_{PQ}=\fr7\d^{(M}_P\d^{N)}_Q
    -\fr{14}f^{A(M}{}_Pf_A{}^{N)}{}_Q
    -\fr{56}\eta^{MN}\eta_{PQ}\komma\cr
&P_{({\bf27000})}^{MN}{}_{PQ}=\Fr67\d^{(M}_P\d^{N)}_Q
    +\fr{14}f^{A(M}{}_Pf_A{}^{N)}{}_Q
    +\Fr3{217}\eta^{MN}\eta_{PQ}\komma\cr
&P_{({\bf248})}^{MN}{}_{PQ}=-\fr{60}f_A{}^{MN}f^A{}_{PQ}\komma\cr
&P_{({\bf30380})}^{MN}{}_{PQ}=\d^{MN}_{PQ}+\fr{60}f_A{}^{MN}f^A{}_{PQ}\komma\cr
}\Eqn\ProjectionOperators
$$
where the structure constants are normalised so that
$f^{MAB}f_{NAB}=-60\d^M_N$.


\acknowledgements
The authors would like to thank Diego Marqu\'es, Jakob Palmkvist and
Henning Samtleben for useful discussions.

\refout

\end